\documentclass[a4paper,11pt]{article}
\usepackage{pos}
\usepackage[utf8]{inputenc}
\usepackage{graphicx}
\usepackage{comment}
\usepackage{amssymb}
\usepackage{amsmath}
\usepackage{wrapfig}
\usepackage{url}
\usepackage{float}

\title{Gamma rays: propagation and detection}
\author*[a]{Elisa Prandini}
\author[b,c]{Konstantinos Dialektopoulos}
\author[d]{Jelena Strišković}

\affiliation[a]{Univerity of Padova,\\
  Via Marzolo 8, 35131, Padova, Italy}
\affiliation[b]{Department of Physics, Nazarbayev University,\\
53 Kabanbay Batyr avenue, 010000 Astana, Kazakhstan}
\affiliation[c]{Laboratory of Physics, Faculty of Engineering, Aristotle University of Thessaloniki, \\
54124 Thessaloniki, Greece}
\affiliation[d]{Josip Juraj Strossmayer University of Osijek, Department of Physics,\\
Trg Ljudevita Gaja 6, 31000 Osijek, Croatia}

\emailAdd{elisa.prandini@unipd.it}
\emailAdd{kdialekt@gmail.com}
\emailAdd{jstriskovic@fizika.unios.hr}

\abstract{Gamma rays constitute a privileged point of view for the study of the extreme Universe. Unlike charged cosmic rays, which are thought to have a common origin, gamma rays are not deflected by galactic and intergalactic magnetic fields. This offers the opportunity to unveil the most powerful particle accelerators, still largely unknown, once modifications in the gamma-ray flux, arrival time, and angular distribution due to propagation effects are considered. 
Gamma ray telescopes include a large variety of instruments, both satellite-born and ground-based, which cover a broad energy range.  These lecture notes provide an overview of the detection techniques for gamma-ray astronomy. A detailed description of the gamma-ray propagation effects in the galactic and extragalactic scenarios is also provided. }

\FullConference{%
  Corfu Summer Institute 2021 ``School and Workshops on Elementary Particle Physics and Gravity''\\
  29 August - 9 October 2021\\
  Corfu, Greece\\
  \vspace{1cm}
  This is an updated version of the manuscript with an improved narrative in Section~2.1.
}

\begin{document}
\maketitle

\section{Introduction: Gamma rays and the Physics of the Universe}

Imagine your eyes suddenly becoming gamma-ray detectors. What would you ``see'' around you? 
The answer is as disappointing as reassuring: you would not see anything other than some rare flashes of gamma-ray light. This happens for two reasons, namely: 1.\ gamma rays are quite rare photons on Earth; and 2.\ astrophysical gamma rays cannot reach Earth due to our natural shield, the atmosphere.

\begin{figure}[ht!]
    \centering
    \includegraphics[width=0.98\textwidth]{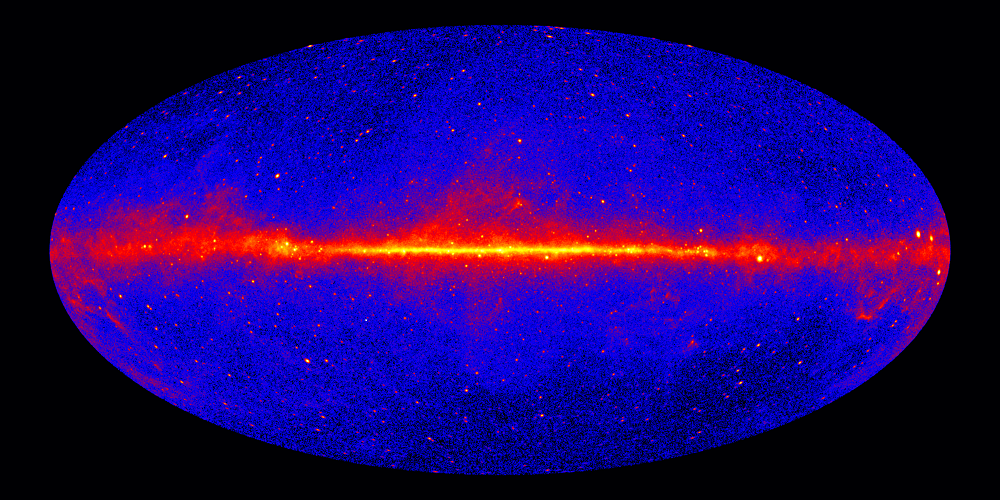}
    \caption{The all-sky map produced by Fermi's Large Area Telescope (LAT), using data from Aug. 4, 2008, to Aug. 4, 2015. Credits: NASA/DOE/Fermi LAT Collaboration.}
    \label{fig:intro_fermi_sky}
\end{figure}

Now imagine that you have this new set of powerful eyes and that you are an astronaut in orbit outside our atmosphere. Then, a wonderful,  bright gamma-ray sky would appear at your sight with a bright diffuse line in correspondence with our Milky Way and thousands of point-like, bright, mostly variable spots filling the whole celestial sphere. This is the sky, represented in Figure~\ref{fig:intro_fermi_sky}, as seen by the {\emph Fermi} satellite, a NASA mission launched in 2008 that has patrolled the gamma-ray sky with no rest since then.

This astonishing image is the result of more than 60 years of technological development, in particular in the field of detector technology. Moreover, this is only a fraction of the vast gamma-ray sky that covers a much larger energy band than that detectable with the instruments onboard the {\emph Fermi} satellite.

This manuscript, which reports the lecture notes of the CA18108 First Training School (Corfu, Greece), will present to the reader an overview of gamma-ray detection techniques and instrumentation. Special emphasis will be put on the capabilities of the different instruments and their limitations.

The second part of the notes is devoted to the propagation of gamma rays. The different phenomena affecting gamma-ray propagation in the Universe will be described. First, the interaction with the extragalactic background light will be treated in detail with several examples from recent experimental results. Then, we will outline other possible effects beyond the standard model affecting this propagation, namely, Lorentz Invariant Violation effects, and axion-like particle oscillation.   
We will treat the subject with a phenomenological approach, underlining the current line of experimental research and the possible future perspectives.

Given the vastness of the subject and the large number of excellent textbooks available covering in detail several aspects of these lectures, these notes are far from exhaustive. The aim is to present an overview of the current and near future gamma-ray astronomical instruments following a historical perspective. 
In this context, gamma-ray propagation effects and their potential to probe cosmology and fundamental physics represent an experimental challenge for current and future experiments. A challenge that the school participants might want to win... 

\begin{figure}[btb]
    \centering
    \includegraphics[width=0.7\textwidth]{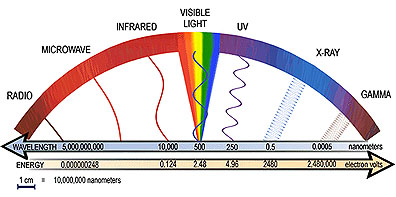}
    \caption{Schematic representation of the electromagnetic spectrum, from \cite{chandra}.}
    \label{fig:intro_em_spectrum}
\end{figure}

\subsection{Gamma-ray astrophysics: nomenclature}
Gamma rays represent the upper end of the electromagnetic spectrum, Figure~\ref{fig:intro_em_spectrum}. 
In astrophysics, gamma rays are defined as photons with an energy larger than 100\,keV. The largest energy photon ever detected to date is a PeV photon recently reported by the LHAASO Collaboration \citep{2021Natur.594...33C}.
This means that the gamma-ray band covers at least 10 orders of magnitude: from $10^5$\,eV to $10^{15}$\,eV and beyond. Hence, the gamma-ray band is the widest electromagnetic band and, as such, the most varied in terms of emission mechanisms and detection techniques.
To have a feeling of the typical quantities we deal with when talking about a gamma-ray photon, we can think that a photon of 1\,TeV energy ($10^{12}$\,eV) has a wavelength of 1.25$\cdot 10^{18}$\,m, a frequency of 2.4$\cdot 10^{-26}$\,Hz and an energy of 1.6$\cdot 10^{-7}$\,J.

Given the wideness of the gamma-ray band, it has been proven useful to define specific intervals (mostly connected with a specific experimental technique). The different bands are listed in Table~\ref{tab:intro_tab_gammaray_bands} together with the main detector mechanism operating in that band and the detector type. The LE (low-energy) band covers the interval between 100 keV and 30\, MeV, this is the typical range of Compton telescopes, usually hosted in balloons or satellites. The HE (high-energy) range between 30\,MeV to 100\,GeV, is instead the regime typical of pair-creation instrumentation onboard satellites. At higher energies the detection is performed from the ground: at the VHE (very-high-energy, 100\,GeV -- 30\,TeV) the typical detectors are atmospheric Cherenkov detectors, at UHU (ultra-high-energy, 30\,TeV--30\,PeV) we find instead water Cherenkov detectors. Finally, at the highest energies (EHE, extreme-high-energy, E $>$ 30\,PeV) the ground-based telescopes use mainly the atmospheric fluorescence technique. 
\begin{table}[h!]
    \centering
    \begin{tabular}{l|c|c|c}
    \hline
    \hline
        Band & Energy & Detector mechanism & Detector type  \\
        \hline
        LE & $<$ 30 MeV & Compton Effect & Balloon/Satellite   \\
        HE & 30 MeV -- 100 GeV & Pair creation & Satellite \\
        VHE & 100 GeV -- 30 TeV & Atm. Chereknov & Ground \\
        UHE & 30 TeV -- 30 PeV & Water Cherenkov & Ground \\
        EHE & $>$ 30 PeV & Atm. fluorecence & Ground  \\
        \hline
        \hline
        \end{tabular}
    \caption{Gamma-ray energy bands and main detection techniques.}
    \label{tab:intro_tab_gammaray_bands}
\end{table}

\subsection{Useful graphs}
When dealing with experimental gamma-ray astrophysics, it might be useful to get familiar with the quantities adopted to study gamma-ray emission from sources.  These are:
\begin{description}
\item[Differential energy spectrum dN/dE:] it is defined as the number of photons per unit of energy, area, and time, usually measured in $1/erg/cm^2/s$ and plotted as a function of energy. The area (and time) is an \emph{effective area (time)} and is related to the detector surface (exposure time) after quality cuts are applied to the data.
\item[Spectral energy distribution $E^2dN/dE$ (or $\nu F_{\nu}$):] is the differential energy multiplied by $E^2$, and when plotted against the energy (or frequency), is referred to as the spectral energy distribution. This representation is very useful when dealing with multiwavelength data because it gives us the opportunity to estimate the total energy carried by photons with different energies. Typical units are $erg/cm^2/s$.
\item[Integral photon flux (light curve):] is the overall emission above a certain energy (threshold), measured in $1/cm^2/s$. It is usually plotted as a function of time and shows the changes of emission over time. This is a key tool in the study of transient or variable phenomena.
\end{description}

To build these plots, the three essential ingredients that strongly affect the performance of a detector are the reconstruction of the energy, the incoming direction, and the arrival time of the gamma-ray photons.

\subsection{Gamma-ray connections: roadmap}

Just as in the case of cosmic rays, the main sources of gamma rays are nearby objects (namely, the Sun), galactic objects, and extragalactic objects.
The study of gamma-ray emission from these sources gives us the possibility of investigating several astrophysical phenomena such as:
\begin{itemize}
    \item Study of jets at galactic scales (microquasars, pulsars), and at extragalactic scales (blazars, radio galaxies, Narrow Lyne Seyfert 1 galaxies);
    \item Interaction between particles and matter, from small scale, e.g., in supernova remnants, to large scale in galaxy clusters;
    \item Transients events from local, solar flares, to galactic magnetars, up to extragalactic fast radio burst and the extremes of the gamma-ray Universe with gamma-ray bursts (GRB).
\end{itemize}

In addition to astrophysics, the study of gamma rays and their absorption has opened a new window for cosmology. Finally, in the past 20 years, a number of exciting studies have proposed the use of gamma rays from astrophysical sources to test the theories of quantum gravity and axion-like particles.  We will come back to this concept in the second part of these lectures.

\subsection{Basic principles of gamma-ray detection}

The essential ingredient to start understanding the basis of gamma-ray detection is the transparency of our atmosphere. As mentioned above, our atmosphere acts as a shell and protects us from most of the high-energy radiation coming from outer space. In addition to charged cosmic rays, extraterrestrial photons are mostly absorbed in the atmosphere. The only exceptions to this rule are radio waves with wavelengths between a few cm and 10\,m, and the tiny optical window. This implies that a direct detection of gamma rays is affordable only in high atmosphere (with balloons) or in outer space, with satellites.
Remarkably, for photon energies in the VHE range on ($>$100 GeV), the passage of a gamma ray in the atmosphere is so ``disturbing'' that the effects can be recorded from the ground.  Therefore, indirect detection of gamma rays has become accessible for ground-based instruments that use our atmosphere as part of their detector. 

\begin{figure}[htb]
    \centering
    \includegraphics[width=0.8\textwidth]{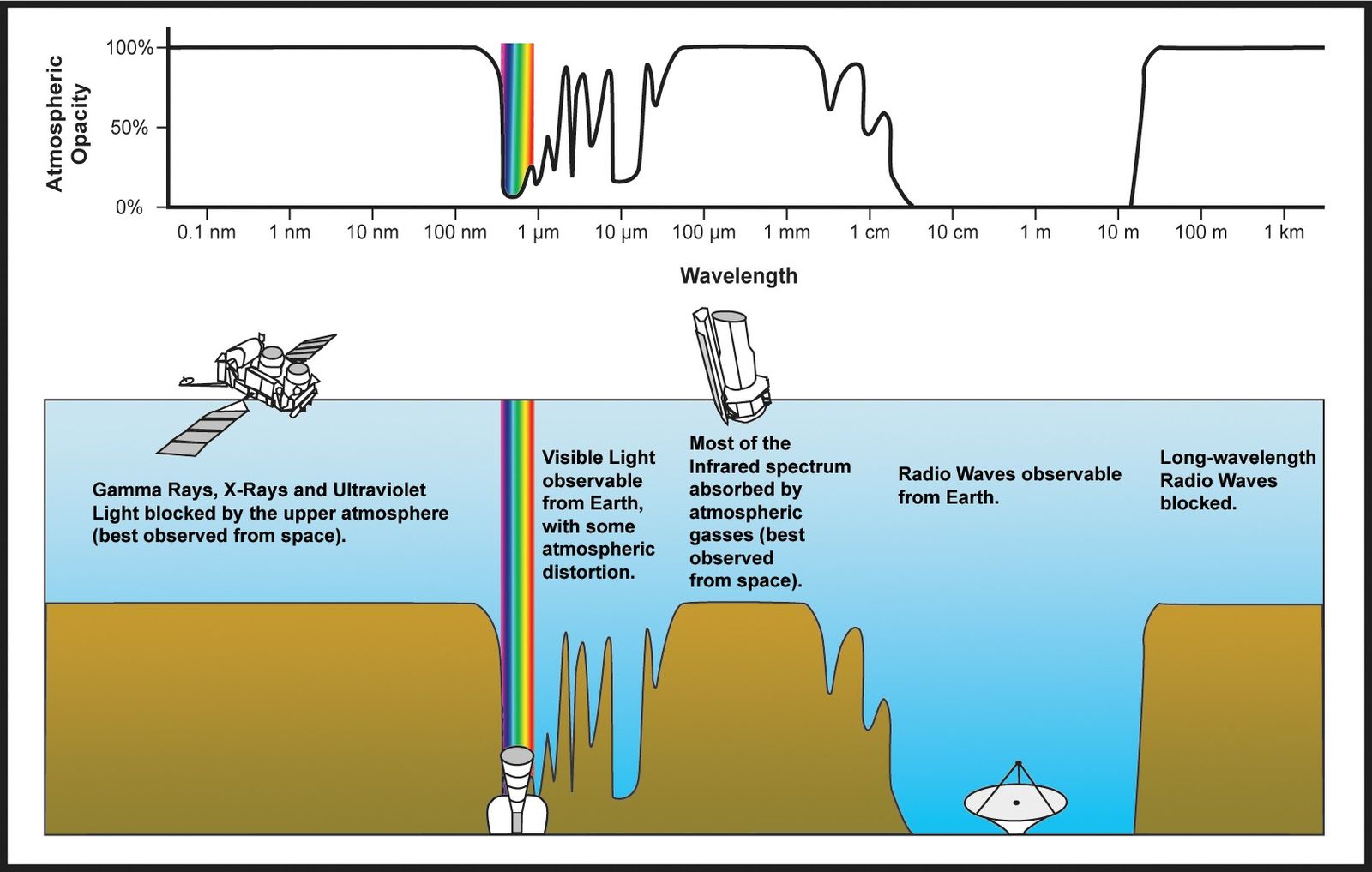}
    \caption{Electromagnetic transmittance, or opacity, of the Earth's atmosphere. Image Credits: NASA.}
    \label{fig:intro_atmospheric_opacity}
\end{figure}

Imagine that you are asked to design a new gamma-ray detector; the first thing that you will probably do is that of studying the performances of existing detectors, placing them in relation to the detector properties, and trying to improve them. In general, the measurement consists in identifying a large number of gamma-ray photons, determining their arrival time, direction, and energy with optimal timing, spatial, and energy resolution. In the most general case, you want to patrol a large portion of the sky and do that non-stop for 24/7.

However, nature is as beautiful as complex and does not easily reveal her secrets. In general, it is impossible for a detector to have all these properties at the same time, and we have to compromise or use a combination of several detectors, each of them ideal for a specific scientific measurement. 

\begin{figure}[htb]
    \centering
    \includegraphics[width=0.9\textwidth]{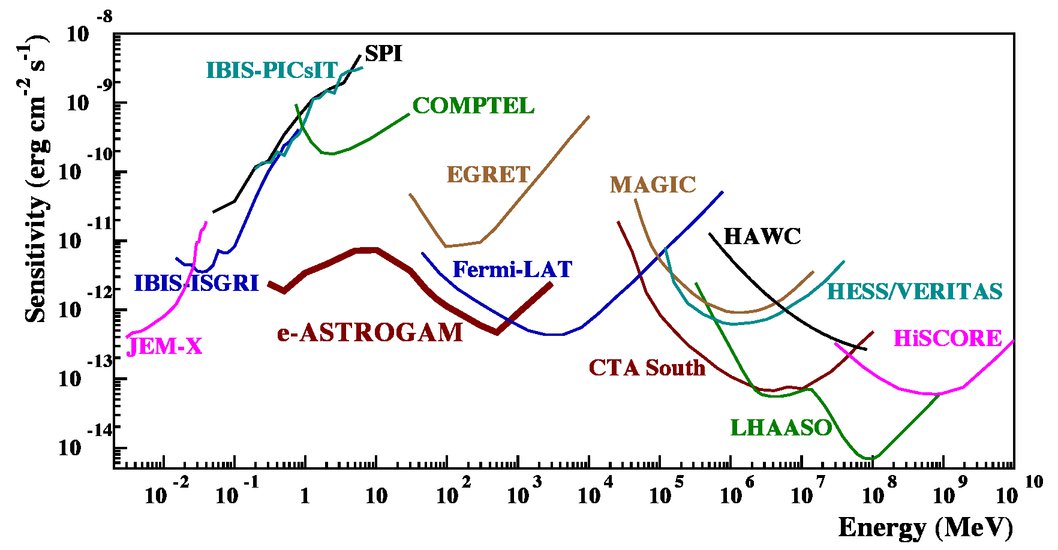}
    \caption{Differential sensitivity for different instruments and exposure times. Point source continuum differential sensitivity of different X- and gamma-ray instruments. The curves for JEM-X, IBIS (ISGRI and PICsIT), and SPI are for an effective observation time Tobs = 1 yr, which is the approximate exposure of the Galactic center region accumulated by INTEGRAL since the beginning of the mission. The COMPTEL and EGRET sensitivities are given for the typical observation time accumulated during the 9 years of the CGRO mission. The Fermi-LAT sensitivity is for a high Galactic latitude source in 10 years of observation in survey mode. For MAGIC, VERITAS (sensitivity of H.E.S.S. is similar), and CTA, the sensitivities are given for Tobs = 50 hours. For HAWC Tobs = 5 yr, for LHAASO Tobs = 1 yr, and for HiSCORE Tobs = 100 h. The e-ASTROGAM sensitivity is calculated at 3$\sigma$ for an effective exposure of 1 year and for a source at high Galactic latitude. From \cite{2018JHEAp..19....1D}.}
    \label{fig:intro_sensitivity}
\end{figure}

\begin{figure}
     \begin{minipage}[c]{0.67\textwidth}
    \includegraphics[width=0.99\textwidth]{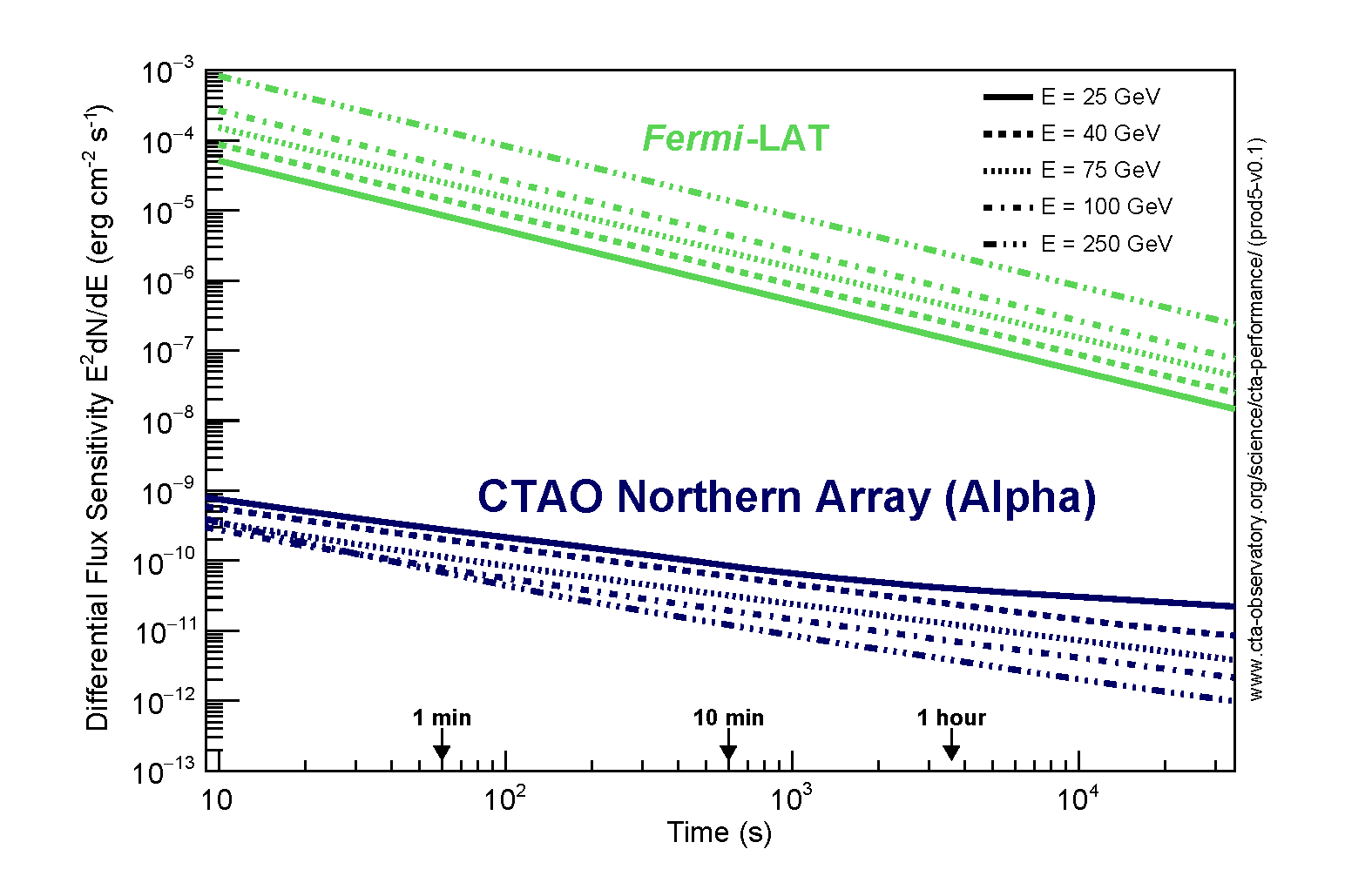}
    \end{minipage} \hfill
    \begin{minipage}[c]{0.3\textwidth}
    \caption{Differential energy sensitivity as a function of time for a ground-based (CTAO) and a satellite-born ({\textit{Fermi}/LAT}) detector. From \cite{cta-observatory}. }
    \label{fig:intro_sensitivity_time}
    \end{minipage}
\end{figure}

\begin{figure}[htb]
    \centering
    \includegraphics[width=0.49\textwidth]{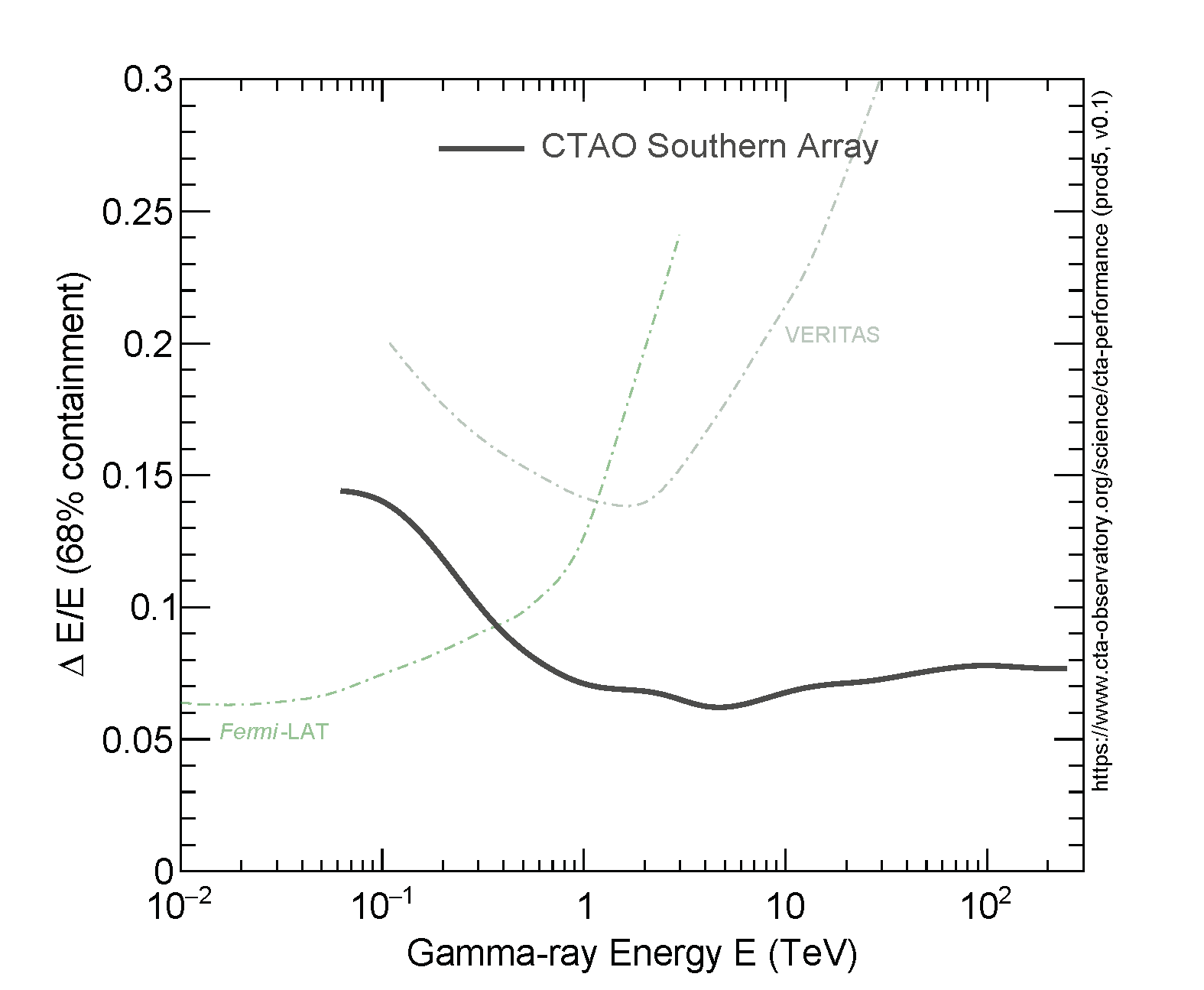}
    \includegraphics[width=0.49\textwidth]{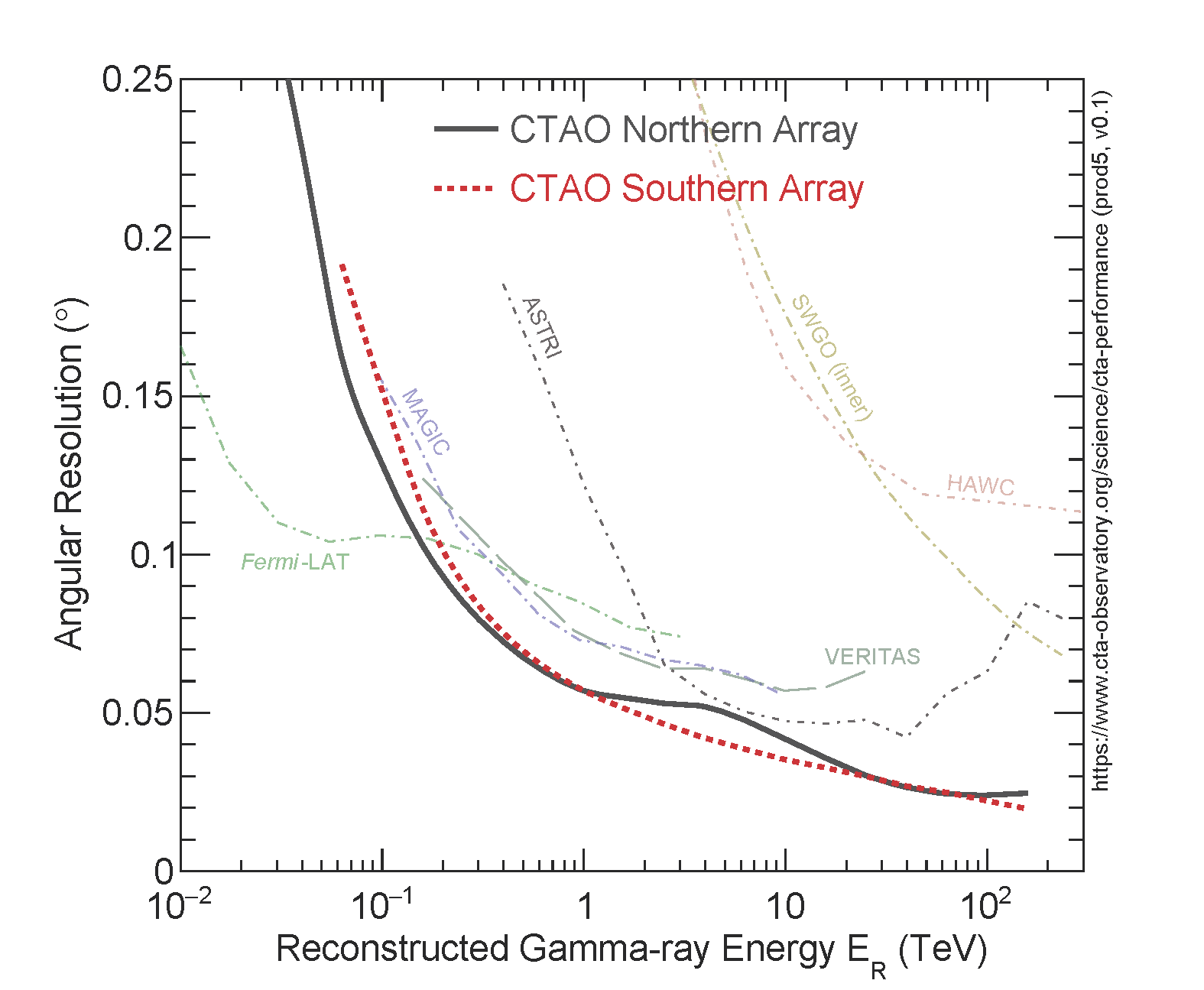}
    \caption{Examples of energy and angular resolution of a gamma-ray instrument. From \cite{cta-observatory}.}
    \label{fig:intro_energy_angular_res}
\end{figure}

The main characteristics used to describe the performance of a detector are:
\begin{description}
\item[Sensitivity:] it is the minimum energy density that a specific instrument can detect in a given amount of time; see Figure~\ref{fig:intro_sensitivity}. The standard is 1 year of exposure time for a satellite and 50 hours for ground-based experiments. 
\item[Transient sensitivity:] is defined as the minimum energy density as a function of the exposure time, Figure~\ref{fig:intro_sensitivity_time}. This is particularly relevant in the case of transient events.
\item[Duty cycle:] is the effective time, along the year, in which a detector observes. It varies from 1000 h/year for a detector that can operate only during moonless nights (e.g., Atmospheric Cherenkov telescopes), to more than 9000 h/year for a detector that is close to a 100\% duty cycle.
\item[Field of view:] it is the portion of the sky (in square degrees or steradians) that can be covered simultaneously by the instrument.
\item[Energy resolution:] is the ability of the instrument to resolve the photon energy. Usually expressed in \%, as $\Delta E/E$ as illustrated in Figure~\ref{fig:intro_energy_angular_res}, left.
\item[Angular resolution:] it is the ability of the instrument to resolve the direction of the incoming photon, expressed in degrees (Figure~\ref{fig:intro_energy_angular_res}, right). In general, the angular resolution of gamma-ray instruments is modest with respect to optical or radio instruments. 
\end{description}

An \textit{ideal detector} would simultaneously have excellent sensitivity, angular, and energy resolution, in addition to a large field of view and a $\sim$100\% duty cycle. This detector does not exist (yet). For example, instruments with a large field-of-view are usually limited by a poor angular and/or energy resolution. For this reason, the identification of the main physics drivers is crucial when planning observations. If we need to locate the emitting region with high precision, our main driver will be an optimal angular resolution, at the expense of other observables, such as the energy resolution. Therefore, when planning observations, we should always identify the instrument that best suits our needs. In case of multiple needs, we might organize a multi-instrument campaign coordinating requests and observations. In the last decade, this has become the standard in high-energy astrophysics, and multiwavelength campaigns with several instruments participating are often organized to fully characterize a specific source or phenomenon in the sky. 

In the case of transient phenomena, the reaction should be fast, and special channels are set up for fast communication. In the field of gamma-ray astronomy, the most common is the Gamma-ray Coordinate Network (GCN)\cite{GCNetwork}.

\section{Gamma-ray detection from satellite}

We have seen that detecting gamma rays means estimating their incoming direction, energy, and time of arrival. However, gamma rays, X-rays, and ultraviolet light are blocked by the upper atmosphere, so it is necessary to go to space if we want to observe them directly. 
In this section, we discuss the attempts to directly observe gamma rays, providing an overview of historical experiments. In addition, we describe the necessary instrumentation that operates in different energy bands. 

\subsection{Milestones of gamma-ray detection from space}
\label{sec:missions}

\begin{figure}[htb]
      \centering
      \includegraphics[scale=0.5]{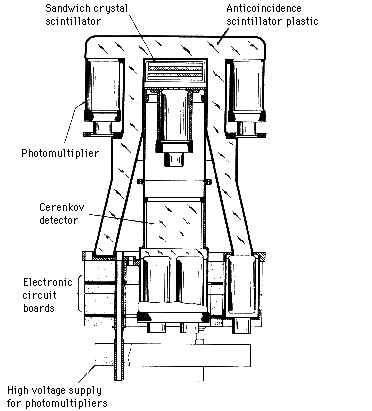}
      \caption{Diagram of the gamma-ray detector of Explorer 11.}
    \label{fig:explorer-11}
\end{figure}

Almost twenty years after gamma rays from extraterrestrial sources were postulated by theoreticians, the space-born gamma ray detection history started. In April 1961 the \textit{Explorer 11} \cite{1965ApJ...141..845K} satellite was launched hosting an instrument designed to detect gamma rays above 50\,MeV, as seen in Figure~\ref{fig:explorer-11}. The detector consisted of a crystal scintillator and a Cerenkov counter, surrounded by a plastic anticoincidence scintillator. It was able to detect 100 photons, but no specific source was identified. 

A few years later the first gamma-ray source was detected: it was our Sun (solar flares).  In the 1970s, NASA's \textit{Second Small Astronomy Satellite} (\textit{SAS-2}) \cite{1975ApJ...198..163F}, together with ESA's mission \textit{COS-B} \cite{1975SSI.....1..245B} detected a number of gamma-ray events, and a first gamma-ray catalog was created (the 2CG catalog \cite{1981ApJ...243L..69S}). However, the angular resolution was poor. In the same years, military defense satellite detectors, which were built to detect flashes of gamma rays from nuclear bomb blasts, recorded the first GRB from deep space. 

\begin{figure}[ht!]
    \begin{center}  
    \includegraphics[width=0.5\textwidth]{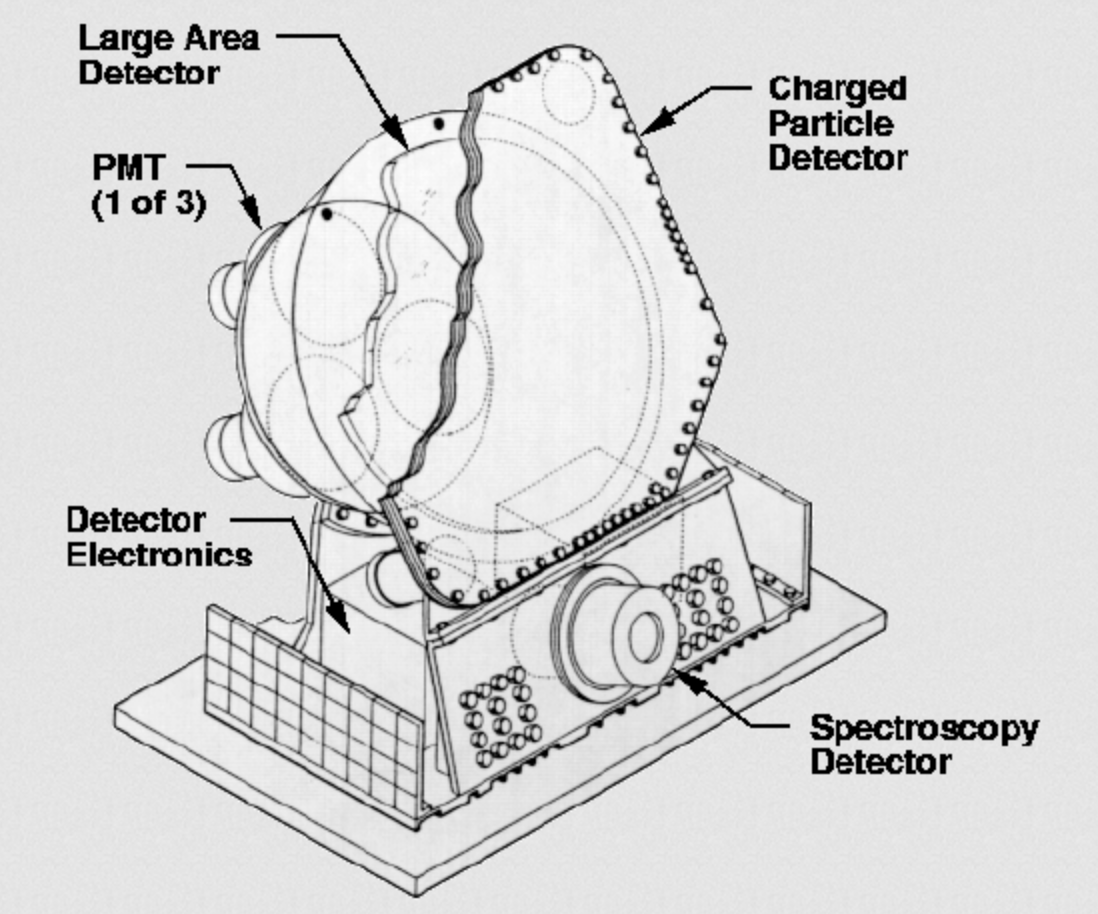}
    \includegraphics[width=0.38\textwidth]{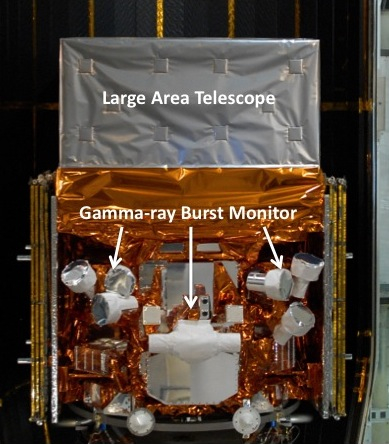}
    \caption{Left image: Scheme of one of the eight modules of the CGRO/BATSE detector \cite{BATSE}. Right image: picture of the \textit{Fermi}/GBM detector \cite{GBM_Fermi}.}    \label{fig:sat_batse_gbm}
    \end{center}
\end{figure}

It is 1987, the supernova 1987A exploded \cite{1987A&A...177L...1W}, emitting gamma rays. In 1991 NASA's mission \textit{Compton Gamma Ray Observatory} (\textit{CGRO}) was launched, Figure~\ref{fig:sat_batse_gbm}, in order to detect photons of energies between 20\,keV and 30\,GeV; that is six orders of magnitude in the electromagnetic spectrum! CGRO carried four instruments: the Burst and Transient Source Experiment (BATSE) \cite{1992NASCP3137...26F}, the Oriented Scintillation Spectrometer Experiment (OSSE) \cite{1992NASCP3137....3C}, the Imaging Compton Telescope (COMPTEL) \cite{1992NASCP3137...85D} and the Energetic Gamma Ray Experiment Telescope (EGRET) \cite{1992NASCP3137..116H} on-board. The first, BATSE, was aimed at GRB detection between 20 and 600\,keV, and it was scanning the sky in its entirety. It managed to detect roughly one event per day in its 9-year duration. OSSE was effective in the 0.05 to 10\,MeV energy range, and was also capable of neutron observations above 10 MeV, mainly for the study of solar flares. It consisted of four detectors, which were normally operated in pairs. The third instrument, COMPTEL, was capable of imaging 1 steradian of the sky between energies of 0.75 and 30\,MeV. It carried two detectors, and gamma rays were detected successively by two interactions: first, a gamma ray would Compton scatter in the upper module and then would be totally absorbed in the lower. In this way, after collecting many events, COMPTEL was capable of creating a map of the position of the sources and an estimate of their photon fluxes and spectra. Finally, EGRET was devoted to energies in the range between 20\,MeV and 30\,GeV, and had a very large field of view. Gamma rays reaching the detector were interacting in a spark chamber, producing positron-electron pairs, and then their tracks and energies were measured within the detector volume. 

Collectively, the CGRO mission had many interesting results: BATSE produced an all-sky map of burst positions, managing to detect approximately 2700 events \cite{batse-exp}. Its results showed definitively that most of the gamma rays were coming from extragalactic objects, and thus a cosmological origin was established. OSSE was able to provide the most comprehensive map of the Galactic center region, revealing the origin of gamma rays from the annihilation of positron-electron pairs \cite{osse-exp}. COMPTEL managed to create an all-sky map of a radioactive isotope of aluminum (${}^{26}$Al), which revealed high concentrations of it in small regions \cite{comptel-exp}. Finally, EGRET produced the first all-sky map of sources above 100\,MeV \cite{egret-exp}. It was capable of detecting 271 events, most of which were unidentified. However, it led to the discovery of blazars.

Later in the 1990s, two high-energy satellites were launched: NASA's \textit{Swift} and ASI/NIVR's \textit{BeppoSAX}. The former, still in operation, consists of three instruments that are meant to detect transient events in the optical, ultraviolet, X-ray, and gamma-ray band. It manages to observe roughly 100 GRBs per year and accurately determine their positions in the sky.  The latter worked efficiently in the energy range between 0.1 and 300\,keV. It operated for six years (1996-2002), and it was able for the first time to quickly detect the position of GRBs on the arc-minute scale and provide X-ray follow-up observations and monitoring. 

In the beginning of 2000, ESA launched its \textit{International Gamma-Ray Astrophysics Laboratory} (\textit{INTEGRAL}) mission \cite{2003AN....324..160W}, which is still operating. It has the ability to detect gamma-ray sources with energy from 15\,keV to 10\,MeV and simultaneously monitor in the X-ray (4-35\,keV) and the optical wavebands. 
  
Late in the 2000s, Italy designed and launched a small satellite called \textit{AGILE} that (still) detects photons both in the X-ray and gamma-ray wavelengths \cite{2008NIMPA.588...52T}. It is a compact instrument ($\sim$60 cm and $<$100 kg) and was designed to operate as a precursor to the \textit{Fermi} satellite. It is sensitive in the 30\,MeV - 50\,GeV energy range.

The most recent mission for gamma ray detection from space is the \textit{Fermi} satellite \cite{2009ApJ...697.1071A}, which was launched by NASA in 2008. It consists of the Large Area Telescope (LAT) and the Gamma-ray Burst Monitor (GBM). The first is an imaging gamma-ray detector operating in the energy range from 20\,MeV to 300\,GeV and has a field of view of about 20\% of the sky. The GBM embodies 14 scintillation detectors and is capable of observing GRB between 150, keV and 30, MeV across the whole of the sky. 

On 2017 August 17th, gravitational waves from a binary neutron star coalescence candidate were observed with the Advanced LIGO and Advanced Virgo detectors \cite{2017ApJ...848L..12A}. The GBM instrument of \textit{Fermi} independently detected a (short) GRB with a $\sim 1.7\,{\rm{s}}$ time delay with respect to the merger time. Together with observations in other bands, Advanced LIGO and Advanced Virgo and GBM observations support the hypothesis that the two events (the merger of the two neutron stars and the GRB) are correlated. 
According to \cite{2017ApJ...848L..12A}, this event was ``the first joint detection of gravitational and electromagnetic radiation from a single source'', a milestone for the field of multi-messenger astronomy. 

\subsection{Basic principles of direct gamma-ray detection}

The standard way to detect photons is to reflect and concentrate them in a confined area of a sensitive detector. However, with an increasing energy of the photon, the difficulty of making it deviate from its initial track increases as well. Therefore, the detection techniques at the base of a canonical ``gamma-ray telescope'' come directly from particle physics detectors. 

A complete treatment of gamma-ray detection, due to its complexity and variegate nature, is far beyond the scope of this lecture. Here, we give an overview of the different techniques adopted in gamma-ray satellites. 

\begin{figure}[htb]
    \centering
    \includegraphics[width=0.7\textwidth]{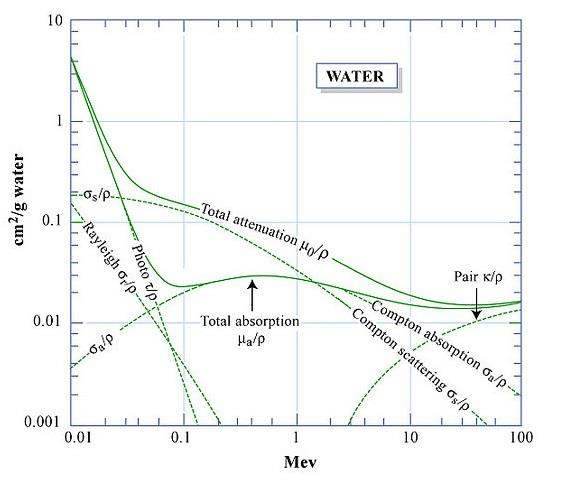}
    \caption{Mass attenuation coefficient for photons in water and relative importance of main interactions for various media. Credits: MIT OpenCourseWare - 22.101 Applied Nuclear Physics, Fall 2006}
    \label{fig:sat_principles}
\end{figure}

Since it is relatively easy to detect a charged particle, an effective method to measure a gamma-ray photon is that of converting (part of) its energy into the energy of electrons released from the medium itself. Depending on the primary photon energy, the three processes that are possible to occur are the photoelectric effect, the Compton effect, and the pair-creation effect. Figure~\ref{fig:sat_principles} shows the mass attenuation coefficient for a photon with an energy ranging from 10 keV to several MeV. Although the photoelectric effect dominates at low energies, around 1 MeV the Compton effect is the predominant process.  
Above a few tens MeV, electron-positron pair production is the main interaction process guiding gamma-ray detection.  

\subsubsection{A generic gamma-ray satellite}
For direct gamma-ray detection, we first need to be able to distinguish gamma rays from the rest of the radiation that comes to our detector. Effectively, this means that we have to get rid of the background radiation, which is mostly composed of charged cosmic rays. 
Therefore, a generic gamma-ray satellite should have a background detector, which is in the form of an \textbf{anti-coincident (AC) shield} that surrounds the gamma-ray detector. 
The main function of this shield is to identify the passage of charged particles (protons, alpha particles, electrons, etc.). Since we are interested in identifying photons, all charged particles are background for us, and their signal in the detector should be rejected. This method is widely used also in high-energy physics, experimental nuclear physics, and particle physics. 

The \textbf{main detector} is aimed at identifying gamma rays and possibly reconstructing their energy and incoming direction. Usually, gamma-ray telescopes host two or more such detectors, with complementary specifications, aimed at different scientific purposes. Photons entering the detector are converted into an electric pulse, which is then detected and recorded by the electronic system of the telescope. The background signals are rejected and the ``good'' photons are tagged by their energy, arrival time and type of event. 

For the whole satellite to work, a \textbf{power supply} is necessary. This is why all such telescopes carry solar panels to ensure energetic autonomy once they are set in orbit. 

Finally, this whole effort would be in vain if the data produced by the detector were not transmitted to the ground. Thus, a \textbf{data transmission system}, which means one or more antennas, is mounted on the satellite for daily (or sometimes more often) data transmission.

\subsection{LE gamma rays}

When a photon carries energy less than 1 MeV, Figure~\ref{fig:intro_sensitivity}, it is considered low energy, and it mostly interacts with matter through the photoelectric effect. What happens during the photoelectric effect is that the photon collides with an electron of the medium and transmits all of its energy. Then, the electron travels through the medium with an energy equal to the energy of the initial photon minus the binding energy. In this way, an electric pulse is produced and can eventually be detected, making us capable of inferring the initial energy of the gamma ray.

To detect photons from the low-energy band, scientists use either scintillators or solid-state detectors. These instruments transform the gamma-ray signal into an optical or electronic signal, which is later recorded. We will discuss both in detail in the following sections. 

\subsubsection{Scintillator detectors}

\begin{figure}[ht!]
    \centering
    \includegraphics[scale=0.5]{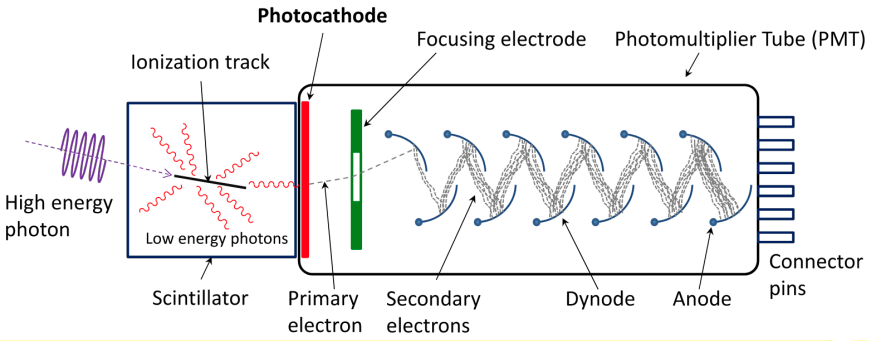}
    \caption{Schematic diagram of a scintillation detector.}
    \label{fig:scintillator}
\end{figure}

To detect fast transients, we have to use instruments that respond extremely quickly, such as scintillators. Specifically,  scintillators are materials in any form, solid, liquid, or gas, that are able to convert high-energy radiation to near-visible or visible light. In gamma-ray detection, the scintillators used are mostly made from inorganic crystals, such as Thallium-doped sodium iodide (NaI(Tl)) or Bismuth gemanate (BGO). A high-energy photon, when passing through a scintillation medium, excites electrons that release low-energy optical/UV photons, which in turn can be easily detected by photomultiplier tubes; see Figure~\ref{fig:scintillator}. 

Scintillators have the great advantage of an extremely fast response (down to ns) and can be easily cut to produce a fine pixelization. However, they are affected by poor energy/spatial resolution. Scintillation detectors can be built with a very large acceptance. These properties (fast response and wide acceptance) make them an ideal instrument for the detection of fast transients. 
Historically, scintillation was the first technique employed for gamma-ray detection. Among scintillator detectors, we mention BATSE, on board the CGRO, and GBM, onboard \textit{Fermi}.

BATSE was a detector composed of eight NaI(Tl) scintillation detectors. Its field of view was $4\pi$\,steradians, and it provided fast triggers for other instruments onboard the CGRO. 
Its successor, the GBM, includes twelve sodium iodide (NaI) scintillators and two cylindrical bismuth germanate (BGO) scintillators.  NaI detectors are sensitive at the lower end of the energy range, from a few keV to about 1\,MeV and provide burst triggers and locations. The BGO detectors cover the energy range ~150\,keV to ~30\,MeV, providing a good overlap with the NaI at the lower end and with the LAT at the high end. The field of view is 9.5\,steradians. After a trigger, the GBM processor calculates the preliminary position and spectral information for telemetry to the ground and possible repointing of \textit{Fermi}. 

\subsubsection{Coded masks}
\begin{figure}[htb]
    \centering
    \includegraphics[width=0.4\textwidth]{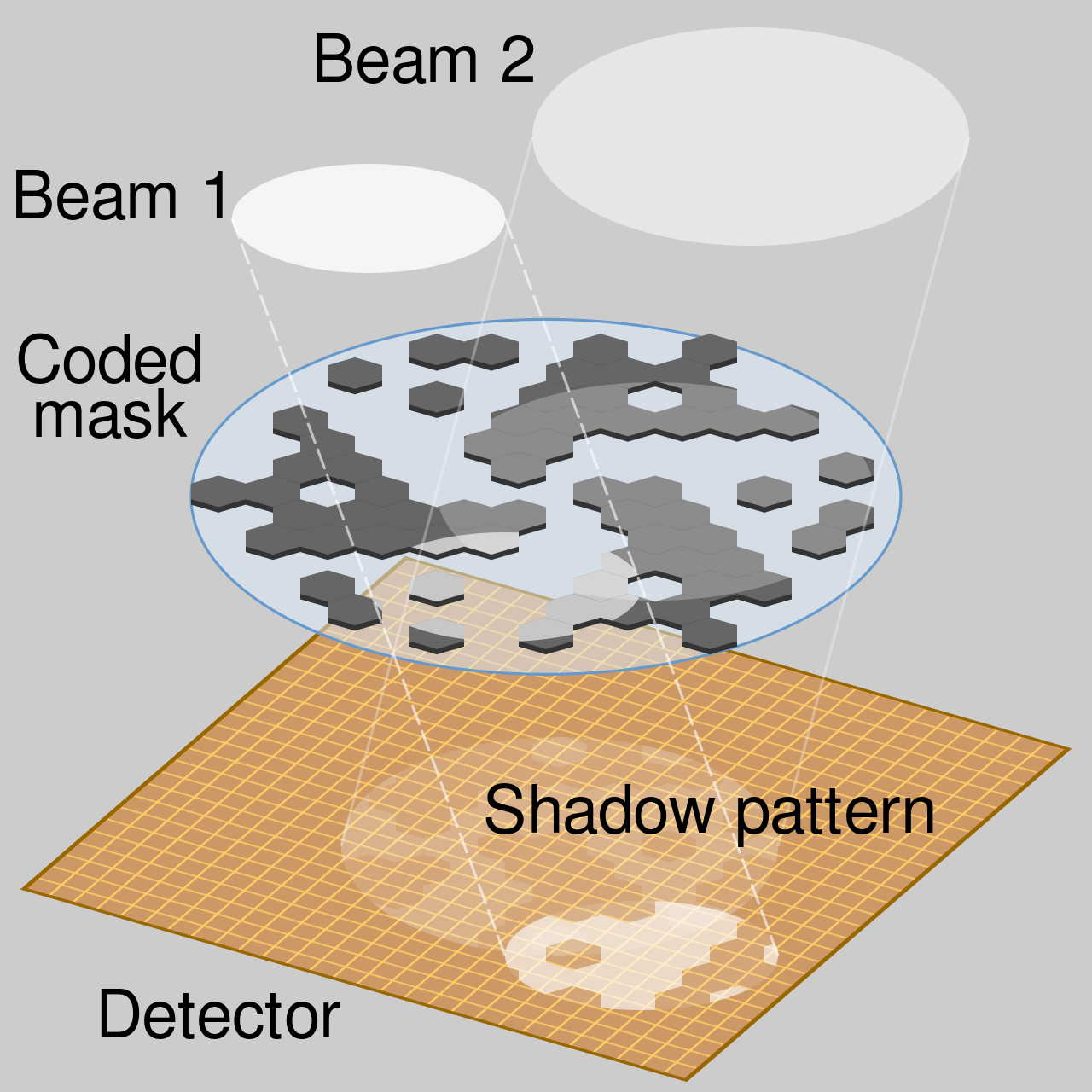}
    \caption{Simplified principle of operation of a coded aperture mask used in the SPI instrument of the INTEGRAL space telescope. Credits: CMG Lee.}    
    \label{fig:sat_codedmask}
\end{figure}

Another method used for the detection of hard X-rays and LE gamma rays is coded mask. A coded mask detector is composed of a mask that filters photons of a certain energy through a specific pattern of opaque material. The surviving photons are then detected via solid-state detectors or scintillators. Reconstruction of the direction and flux of primary photons is done using matrices, as sketched in Figure~\ref{fig:sat_codedmask}.
An example of two different coded mask detectors are the Imager on-Board the INTEGRAL Satellite (IBIS) and the Spectrometer of \textit{INTEGRAL} (SPI). Similar instruments were also used in NASA's \textit{Swift}-BAT, in \textit{BeppoSAX} (Wide Field Camera) and others. Such instruments are less sensitive compared to the ones with AC shields, but can be more useful in Galactic astrophysics, since they have the ability to generate images, thus reducing the problem of source confusion. 

As we already mentioned in Section~\ref{sec:missions}, \textit{INTEGRAL}'s objective is to gather as much information as possible for the most violent and energetic objects in space, and this becomes possible because it combines fine spectroscopy and fine imaging of gamma-ray emissions, while monitoring in X-ray and optical bands as well. 

In its entirety, \textit{INTEGRAL}'s payload module consists of four instruments: the IBIS and SPI mentioned above, together with the Joint European X-ray Monitor (JEM-X) and the Optical Monitoring Camera (OMC). IBIS is capable of observing from 15\,keV to 10\,MeV, having an angular resolution of 12\,arcmin, and thus being able to locate a bright source to better than 1\,arcmin. SPI can detect radiation between 20\,keV and 8\,MeV. These two instruments together with JEM-X, which is responsible for monitoring soft and hard X-ray sources from 3 to 35\,keV, share a common feature: they operate as coded mask telescopes (using rectangular, hexagonal and hexagonal tiles respectively). The OMC is sensitive between 500 and 580\,nm, thus being able to monitor the optical band. 

\subsubsection{Compton telescopes}
\begin{figure*}[h!]
  \begin{minipage}{.5\linewidth}
    \centering
    \includegraphics[width=1\linewidth]{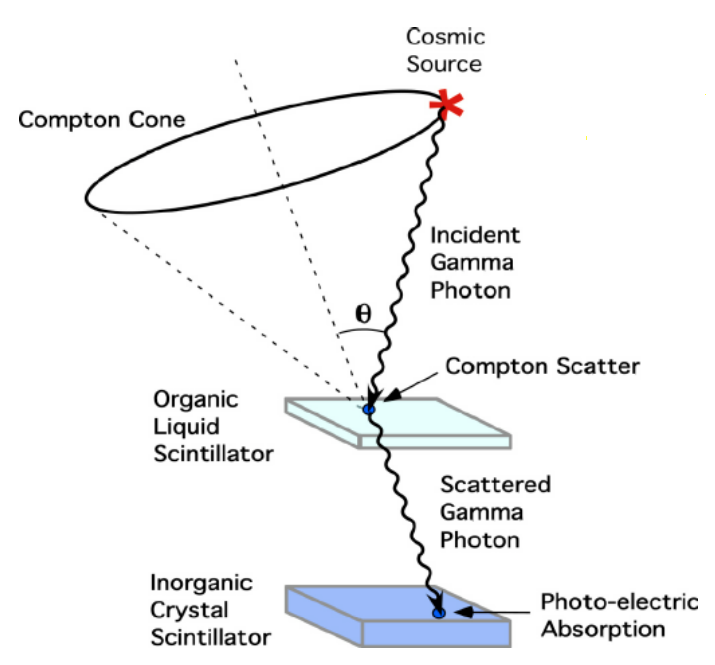}
    \caption{Schematic representation of a Compton detector.}
    \label{fig:Compton}
  \end{minipage}%
  \begin{minipage}{.5\linewidth}
    \centering
    \includegraphics[width=.68\linewidth]{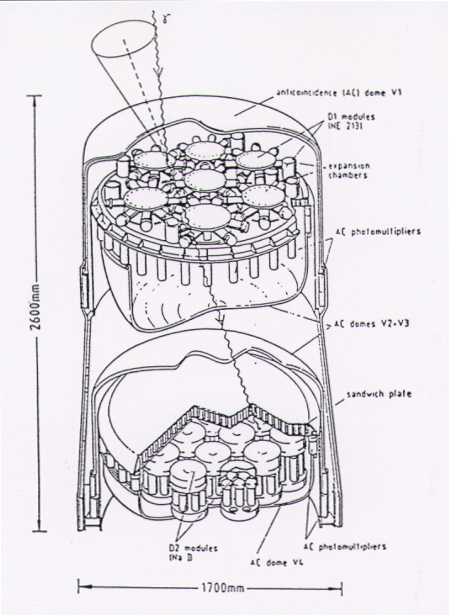}
    \caption{Schematics of COMPTEL.}
    \label{fig:COMPTEL}
  \end{minipage}
\end{figure*}

When the energy of the photon is in the MeV regime, the process that becomes dominant is the Compton effect. 
Compton telescopes have two layers of detection: first, the incident photon is scattered by the initial detector, and then it is absorbed by the second one, as shown in Figure~\ref{fig:Compton}. In this way, if we know that the energies of the incoming photons in the two detectors are, say, $E_1$ and $E_2$, we can immediately calculate the scattering angle from:
\begin{equation}
    \cos \phi = 1 - \frac{m_e c^2}{E_2}+\frac{m_e c^2}{E_1+E_2}\,.
\end{equation}
It is possible to increase angular resolution using coded masks, which are effective at energies below $\sim$1\,MeV. Compton telescopes have a large field of view, about 1 steradian, but their sensitivity compared to instruments in other energy regimes remains low. 

An example of Compton telescopes is COMPTEL, which was on board the CGRO that we discussed earlier in Section~\ref{sec:missions}. The first detector of COMTPEL used a liquid scintillator, while the second one used NaI crystals. Each of these detectors, as can be seen in Figure~\ref{fig:COMPTEL}, was surrounded by an AC shield of plastic scintillator to reject charged particles. 

Among the future proposed missions, \textit{e-ASTROGAM} (enhanced ASTROGAM) concerns this intermediate energy regime. 
The proposed payload consists of:
\begin{itemize}
    \item A tracker, which will utilize two techniques to detect cosmic gamma rays: Compton scattering and pair production. 
    \item A calorimeter to measure the energy of secondary particles. 
    \item An AC shield to reject charged particles from the background. 
\end{itemize}
It is planned to operate in the energy range between 0.3\,MeV and 3\,GeV, and it will improve the sensitivity in the MeV regime up to two orders of magnitude compared to \textit{COMPTEL}. In addition, it will be able to measure the polarization of the radiation in this range, thus contributing to the multi-messenger era.  

\subsection{HE gamma rays: pair conversion telescopes}
As the energy of the photon exceeds several MeV, the pair production process starts to dominate the photon-matter interaction. Hence, between $\sim$100\,MeV to $\sim$100\,GeV, pair conversion telescopes are typical gamma-ray detectors. Apart from the photon energy, another factor that increases the probability of pair production in interactions between photons and matter is the square of the atomic number of the atoms of the medium. This is why high-Z elements are used in gamma-ray detectors to trace electron-positron trajectories.

Specifically, after the creation of a positron-electron pair, the particles penetrate through the detector, allowing the path of the particles to be tracked. At the bottom of the pair conversion telescope, usually a calorimeter is located to measure the energy of the particles. Knowing their trajectories and energy, we can infer the energy and incoming direction of the primary photon. 
\begin{figure}[ht!]
  \begin{minipage}{.33\linewidth}
    \centering
    \includegraphics[scale=0.38]{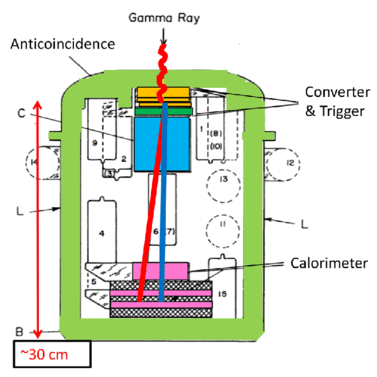}
  \end{minipage}%
  \begin{minipage}{.33\linewidth}
    \centering
    \vspace{0.3cm}
    \includegraphics[scale=0.5]{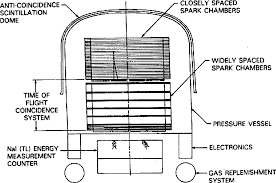}
  \end{minipage}%
  \begin{minipage}{.33\linewidth}
    \centering
    \includegraphics[scale=0.26]{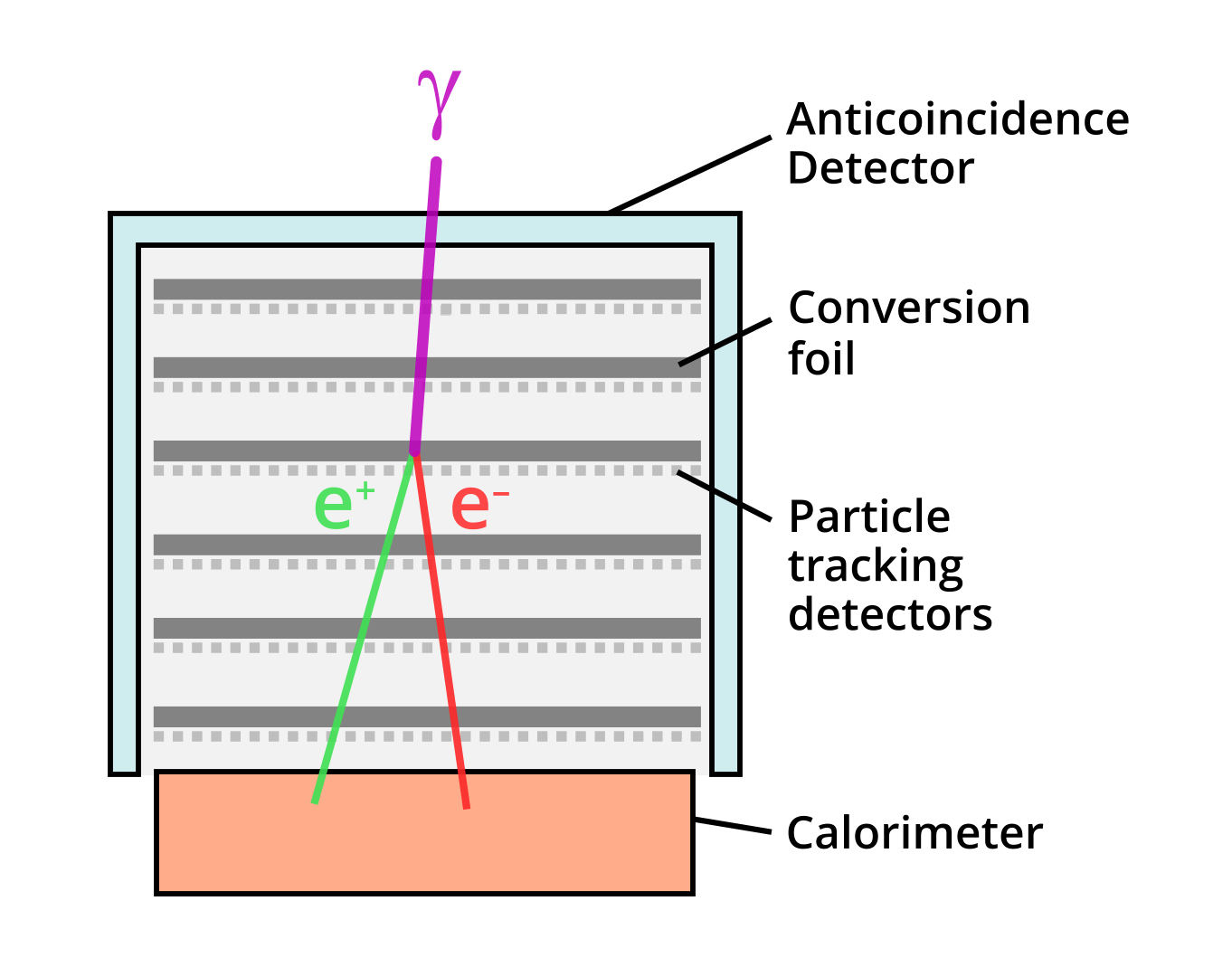}
  \end{minipage}
  \caption{The three different pair conversion telescopes are shown in chronological order: from left to right we see OSO-3, launched in 1967, the EGRET that was on-board CGRO, launched in 1991 and the \textit{Fermi}-LAT,  launched in 2008. All of them were operated by NASA.}
  \label{fig:pair_prod_tele}
\end{figure}

The evolution of pair conversion telescopes over the years can be appreciated in Figure~\ref{fig:pair_prod_tele}. The Orbiting Solar Observatory 3 (OSO-3), launched in 1967 by NASA, used a scintillation crystal of NaI(Tl) together with a phototube, inside an AC shield. It operated in the energy range between 7.7 and 210\,keV. More than two decades later, NASA launched EGRET, in 1991, which was on board the CGRO satellite. This one used a spark chamber to capture and record the high-energy radiation, while it was enclosed in an AC dome to veto the background charged particles. The revolution came with \textit{Fermi}, launched in 2008, planned to operate for up to 10 years, while it still provides us with data. Its working principle is based on silicon detectors that measure the passage of charged particles. As can be seen in Figure~\ref{fig:pair_prod_tele}, the different layers of silicon and the conversion foils help us reconstruct the path traveled by the particles. After passing through these layers (the tracker), charged particles fall onto the calorimeter, which uses scintillator crystals (CsI) and measures their energy. In addition to the tracker and calorimeter, the detector of LAT consists of a segmented AC shield, which utilizes 89 plastic scintillator tiles, and its aim is to measure background radiation. 


\subsection{Summary of gamma-ray space-born detectors}
To summarize, the main properties of a gamma-ray satellite in the LE and HE ranges are listed in Table~\ref{tab:sat_detector_comparison}.

\begin{table}[h!]
    \centering
    \begin{tabular}{l|c|c|c|c|c}
        \hline
        \hline
        Energy  & Detection  & Detector   &  Energy  & PSF  & FoV  \\
         Range &  technique  &  &    resolution          &      &      \\
        \hline
        LE & Coded Mask & INTEGRAL/SPI  & 0.2\% & 2.5$^{\circ}$  & 14$^{\circ}$ flat to flat\\
        LE & Coded Mask & INTEGRAL/IBIS  & 8-10\% &$<$10'& 8.3 x 8$^{\circ}$  \\
        LE & Scintillation &  CGRO/BATSE &  $<$10\% &   25$^{\circ}$ (alert) &  4$\pi$ sr    \\
        LE & Scintillation & Fermi/GBM  &  $<$10\% &   15$^{\circ}$ (alert)   & All sky but Earth     \\
        LE & Compton & Comptel (1991)  & 5-8\% &  1.7 - 4.4$^{\circ}$&    64$^{\circ}$  \\
        LE & Compton &  e-ASTROGAM & 3\% & 1.5$^{\circ}$ & 2.5 sr\\
        HE & Pairs & CGRO/EGRET & 20\% & 5.8$^{\circ}$ & 0.6\,sr \\
        HE & Pairs & Fermi/LAT  & $<$10\% & 0.15$^{\circ}$-3.5$^{\circ}$ & $>$2\,sr\\
    \hline
    \hline
    \end{tabular}
    \caption{Comparison of the main gamma-ray satellite detectors.}
    \label{tab:sat_detector_comparison}
\end{table}

\newpage

\section{Gamma-ray ground-based detection}

\begin{figure}[ht!]
    \centering
    \includegraphics[width=0.75\textwidth]{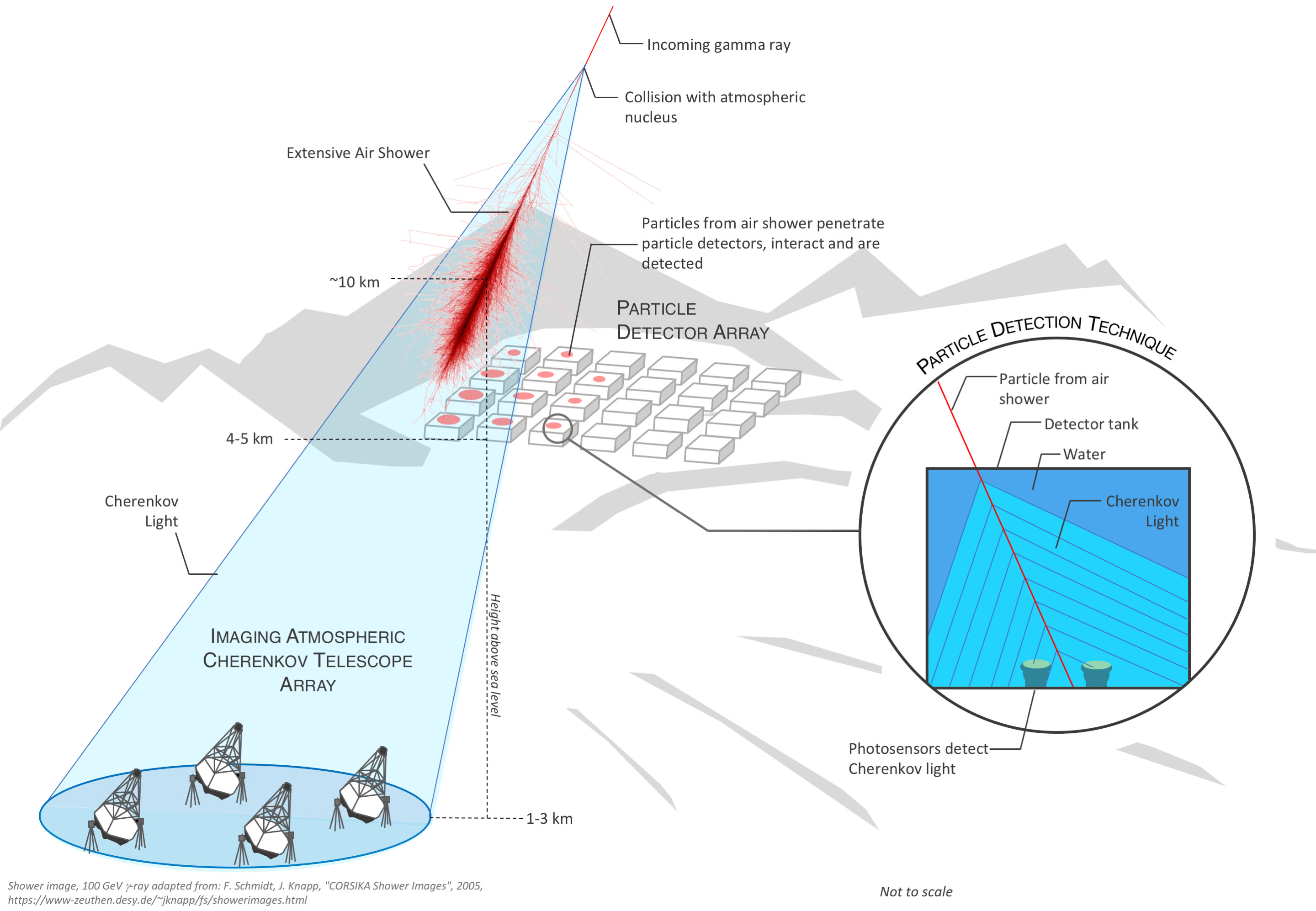}
    \caption{Gamma-ray detection methods from ground based detectors \cite{swgo}.}
    \label{fig:ground_detection_concept}
\end{figure}

Direct gamma-ray detection from the ground is not feasible due to the unavoidable interaction of gamma rays with atmospheric nuclei. However, this interaction initiates a multiplicative process that converts the primary gamma ray into a large number of particles, namely low-energy photons, electrons, and positrons, forming what is called an extensive air shower (EAS). This shower has a maximum at $\sim$10\,km altitude, and particles can be measured directly up to an altitude of a few kilometers. Only secondary radiation composed of ultraviolet-visible (UV-optical) photons known as Cherenkov radiation penetrates the inner layer of the atmosphere reaching the ground. This is represented in Figure~\ref{fig:ground_detection_concept}.

In the following sections, we outline the main properties of EAS, followed by a detailed description of Imaging Atmospheric Cherenkov detectors and EAS detectors. The former, situated at moderate altitudes, are aimed at collecting Cherenkov light from the shower, and the latter, located at higher altitudes, are aimed at directly detecting charged particles.

\subsection{The basis of Extensive Air Shower}
\begin{figure}[th]
    \centering
    \includegraphics[width=0.7\textwidth]{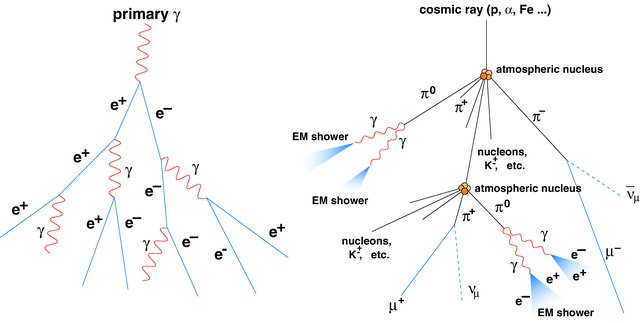}
    \caption{Schematic of air shower development. Electromagnetic shower on the left and hadronic shower on the right. Credit: R.M. Wagner, dissertation, MPI Munich 2007.}
    \label{fig:em_shower}
\end{figure}

After the first interaction with a nucleus, a gamma-ray is converted into an electron-positron pair. Each particle of the pair, in turn, produces a photon via Brehmsstrahlung. This multiplicative process is schematically represented in Figure~\ref{fig:em_shower}, left. Interestingly, a similar but more complex process occurs with protons (and hadrons, in general). The resulting showers are represented in Figure~\ref{fig:em_shower}, right.

A useful toy model to describe the development of electromagnetic showers was proposed by Heitler in 1944 \citep{1954qtr..book.....H}. Under the simplifying hypotheses that (i) the interaction cross section is independent of the particle and its energy; and (ii) the ionization and excitation effects are neglected, and therefore only Brehmsstrahlung is effective, we have the following:
\begin{equation}
    E(X) = E_0e^{-X/X_0}\,,
\end{equation}
where $E_0$ is the energy of the primary particle that initiates the shower and $E$ is its energy. $X$, expressed in g\,cm$^{-2}$, is the product of the traveled path in the atmosphere (in cm) and the atmospheric density $\rho$ that varies with altitude. $X_0$ is the radiation length, that is, the mean free path of an electron with high energy. In the atmosphere of the Earth, we have $X_0 \simeq\,37$\,g\,cm$^{-2}$. According to this model, each step of the cascade has an average length value $d = \rm{ln}2 \cdot X_0$. The shower develops to a maximum. Above this maximum, when the particles have an average energy named critical energy $E_c\simeq86\,MeV$, excitation/ionization losses become important and the development of the shower stops.

The number of particles in step $n$ is $N=2^n$. At maximum, N is $N_{max} = 2^k\sim E_0/E_c$: it is an extremely high number! To get a feeling, estimate $N_{max}$ for a shower initiated with 1\,TeV gamma ray. 
The number of particles in a shower is often referred to as the shower \textit{size}. 

Due to their relativistic energies, all particles of a shower are strongly collimated along the incident direction. The main process that broadens the shower transversely is the multiple scattering and, in the second order, the deflection of Earth's 
magnetic field. Electrons and positrons can also induce electromagnetic showers with characteristics identical to those induced by $\gamma$--rays. Thus, electrons and positrons 
should be considered as an irreducible background for ground--based gamma ray detectors. 

Rossi and Greisen proposed more precise analytical models of the development of these showers in the mid-40s \cite{1941RvMP...13..240R}.
These theories have been developed in the so-called ``B approximation'': every process is neglected except for pair production and energetic loss by Bremsstrahlung. 
The analytical solution for $N_e$ (that is, the total number of electrons and positrons above the critical energy) is:
\begin{equation}
N_e(t,E_0)=\frac{0.31}{\sqrt{\ln(E_0/E_c)}}\cdot e^{t\cdot(1-1.5\ln s)}\,,
\end{equation}
where $t$ is the atmospheric depth, $s$ is the \emph{age parameter}, which indicates the level of development of the shower and goes from a value of 0 at the point of first interaction to 1 at maximum and 2 at the point where the shower dies.

The variation of $N_e$ with $t$ is often called \emph{longitudinal development} of the shower. The \emph{lateral distribution} of a shower has the main contribution of low-energy particles, i.e., those produced at lower altitudes.

\begin{figure}[htb]
    \centering
    \includegraphics[width=0.7\textwidth]{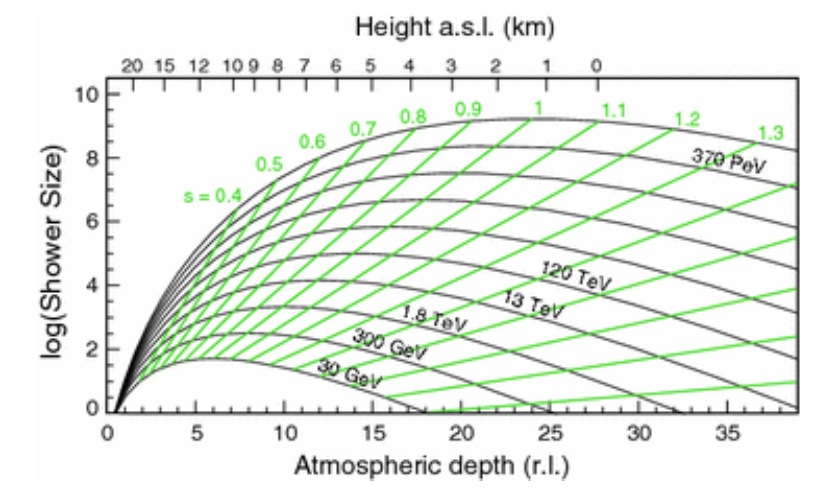}
    \caption{Longitudinal shower development from a photon-initiated cascade. Credit: R.M. Wagner, dissertation, MPI Munich 2007.}
    \label{fig:showers_s}
\end{figure}
Figure~\ref{fig:showers_s} illustrates the dependence of the shower \textit{size} on the atmospheric depth in units of radiation length r.l. (and altitude a.s.l. in km, top axis) for showers initiated by gamma rays of energy ranging from 30\,GeV to 370\,PeV. A 30\,GeV shower reaches its maximum at $\sim 6$ r.l., 12\,km a.s.l., while a shower initiated by a gamma ray of 370\,PeV reaches its maximum at an altitude of roughly 7\,km. 

For direct detection of a gamma-ray shower, a shower $size$ $\geq$\,100 is needed. Therefore, the energy threshold of an EAS detector is strongly dependent on the altitude of the observatory. The best feasible case is a detector placed at 4 to 5\,km altitude. We expect, with standard analysis techniques, an energy threshold of $\sim$1\,TeV that is the threshold of the current generation of EAS detectors. 
To go below this threshold, an alternative technique has been developed: the Imaging Atmospheric Cherenkov Technique (IACT), described in the next Section. 

\subsection{Indirect detection of EAS: IACTs}
 
The IACT technique works between tens of GeV and tens of TeV. It is based on the detection of Cherenkov light emitted by atmospheric showers induced by cosmic rays. The Cherenkov effect occurs when a charged particle travels in a dielectric of refractive index \emph{n} with a speed exceeding the speed of light in the medium ($c/n$). When a charged particle moves in a dielectric medium, like air, polarization occurs. When the velocity of the particle v is \emph{superluminal}\footnote{In a dielectric medium with refractive index \emph{n}, the motion of a particle is superluminal when its velocity is higher than the light speed in that medium $c/n$.}, the particle is moving faster than the electromagnetic wave induced by the polarization; in this case, a coherent wave front appears at an angle $\theta$, and the radiation emitted is called \emph{Cherenkov light}.

An analytical description of the Cherenkov effect can be given using a classical approach. When the particle moves in the medium, it looks like it is emitting a succession of spherical waves that propagate with velocity $c$. If the motion of the particle is superluminal, a coherent front wave appears, separating the external area, which has no signal, from the internal one. In the internal region, the waves are superimposed, meaning that at every point of this region, two delayed electron positions coexist and an electric field is formed. 

The angle of emission can be derived with simple geometrical arguments, yielding the following:
\begin{equation}\label{eq:cosTheta}
\cos\theta=\frac{1}{\beta n}\,,
\end{equation}
 where $\beta=v/c$. The surface delimited by this angle is called
 \emph{Cherenkov cone}.  The limit Cherenkov angle is $\theta_{max}= arccos(n^{-1})$, which occurs if $\beta=1$ (ultrarelativistic particles). As already mentioned, the threshold velocity for Cherenkov light emission
is 
\begin{equation}
\beta_{min}=1/n\,,
\end{equation}
which implies a minimum energy threshold for particles of:
\begin{equation}\label{eq:Emin}
E_{th}=\frac{m_0c^2}{\sqrt{1-\beta^2_{min}}}=\frac{m_0c^2}{\sqrt{1-n^{-2}}}\,,
\end{equation}
where $m_0$ is the rest mass of the particle.
At sea level, the air refractive index is $\sim1.00029$.  This corresponds to an energy threshold of 21\,MeV for electrons, 4.3\,GeV for muons, and 39\,GeV for protons.

The number of emitted photons as a function of the path length and wavelength is as follows:
\begin{equation}
\frac{dN}{dld\lambda}=2\pi\alpha\left(1-\frac{1}{(\beta n)^2}\right)\cdot\frac{1}{\lambda^2}\,,
\end{equation}
where $\alpha$ is the fine structure constant.

The strength lines of the electric field are radial and point to the actual position of the emitting particle. The electric field diverges on the Cherenkov cone surface, which is a singular surface where the sharp transition of the electric field occurs. This front wave propagates at the speed of light $c$, forming a kind of electromagnetic collision wave that is at the origin of the impulsive signal registered by the detectors. Cherenkov radiation can occur only at frequencies for which $n > 1$, i.e., from microwave to UV, but not at higher energies of the electromagnetic spectrum.
 
The low energy threshold for Cherenkov light emission of electrons (Eq.~\ref{eq:Emin}) makes them the main emitters of Cherenkov light in atmospheric showers. Electrons in these showers emit Cherenkov radiation at different altitudes. Thus, the correct model for describing the Cherenkov light emitted by electrons should take into consideration atmospheric variation with respect to the altitude. 

We assume that the atmosphere density $\rho$ scales exponentially with altitude h (the so--called isothermal atmosphere approximation):
\begin{equation}
\rho(h)=\rho_0\cdot\exp(-h/h_0)\,,
\end{equation}
where $h_0=7.1$\,km is the scale--height, and $\rho_0=0.0013\,$g/cm$^3$ is the air density at sea level.
In this approximation, the refractive index is then:
\begin{equation}
n=1+\eta_h=1+\eta_0\exp(-h/h_0)=1+2.9\cdot10^{-4}\exp(-h/h_0)
\end{equation}
The dependence of the refractive index of the wavelength $\lambda$ can be neglected for our values of $\lambda$.

The threshold energy for electrons according to Eq. \eqref{eq:Emin} is:
\begin{equation}
E_{th}(h)=\frac{0.511\,\mathrm{ MeV}}{\sqrt{2\eta_0\exp(-h/h_0)}}\,,
\end{equation}
The angle $\theta_{max}$ can be written as:
\begin{equation}
\cos\theta_{max}(h)=\frac{1}{1+\eta_h}\simeq1-\eta_0\exp(-h/h_0)\,.
\end{equation}

\begin{figure}[ht]
\centering
\includegraphics[width=0.4\linewidth]{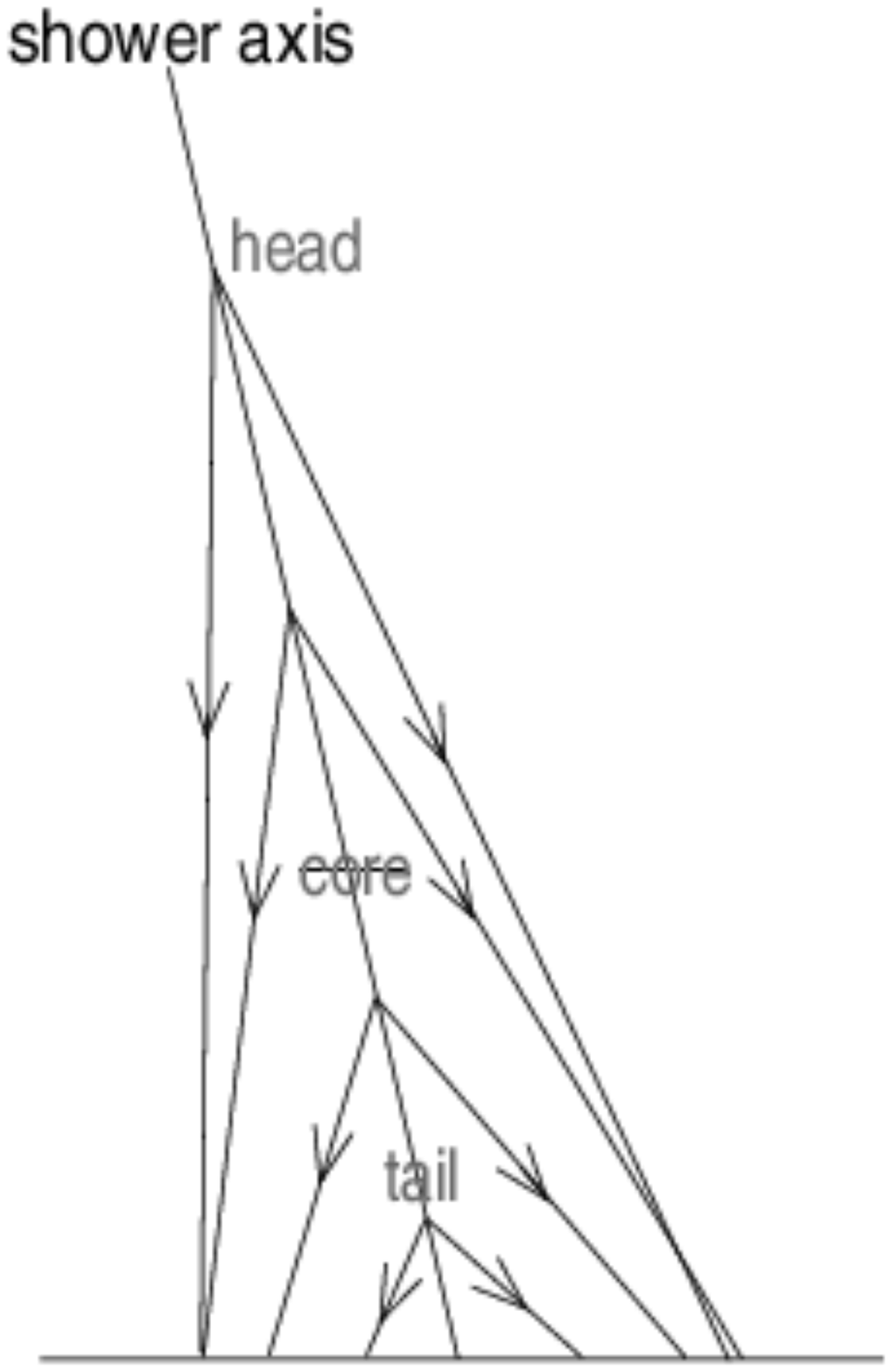}
\caption{Schematic view of the emission Cherenkov angle at different altitudes of an atmospheric shower.}
\label{img:CHERshower}
\end{figure}

In Figure~\ref{img:CHERshower}, the Cherenkov emission angle at different altitudes for an atmospheric shower is sketched. The radius has a maximum, corresponding to photons emitted between 10 and 20\,km. The light from the shower tail is emitted with a larger Cherenkov angle, but its distance is smaller due to the lower height.

\begin{figure}[htb]
    \centering
    \includegraphics[width=0.9\textwidth]{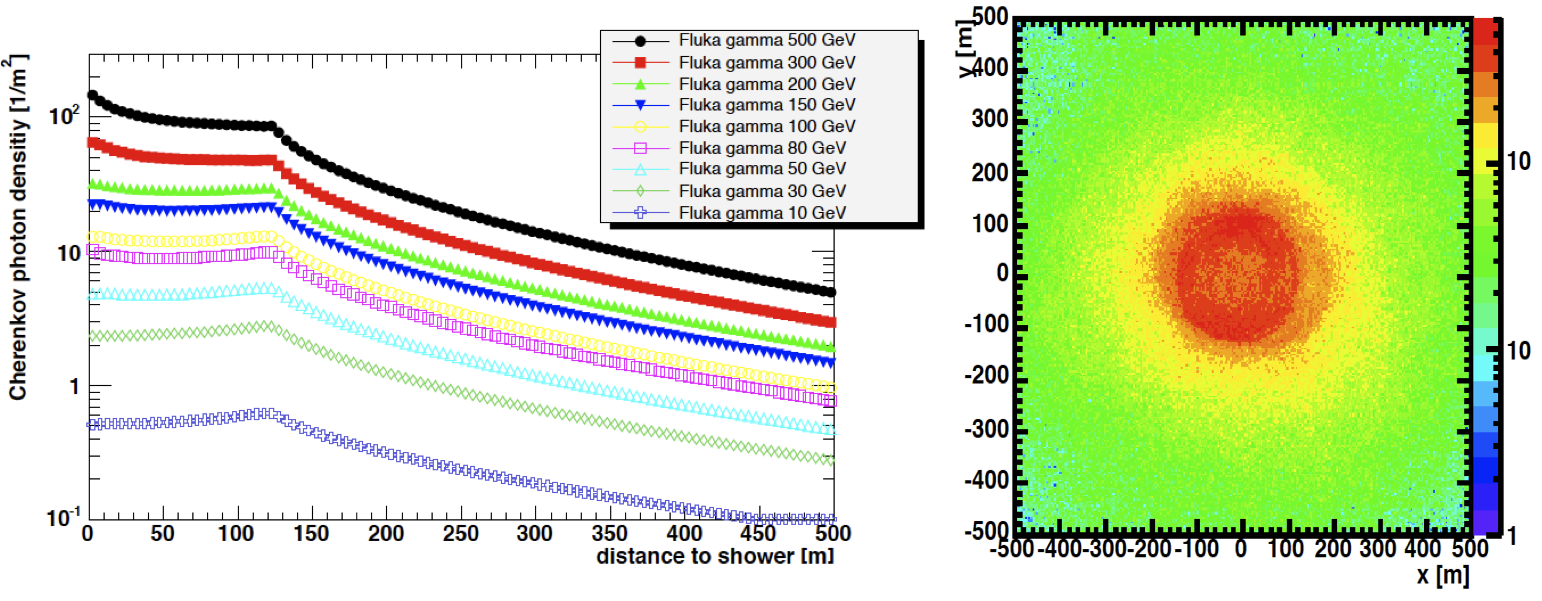}
    \caption{Monte Carlo simulations of the distribution of Cherenkov photons on the ground for gamma-ray initiated air showers. The left plot shows the Cherenkov photon density as a function of radial distance from the shower core for primaries with a range of energies, the right shows the two-dimensional photon density on the ground for a shower with a 300\,GeV primary energy. Figure courtesy of G. Maier, from \cite{2015arXiv151005675H}.}
    \label{fig:cherenkov_light_pool}
\end{figure}

Usually, Cherenkov photons spread over a quite large area, that is a circle with a diameter of $\sim 300$\,m, for altitudes of 2200\,m, Figure~\ref{fig:cherenkov_light_pool}.  This huge surface allows the detection of showers with a large \textit{Impact Parameter} (IP: distance between the shower axis and the point of detection). However, this large area decreases the photon density.

The \emph{zenith angle} of the observation is another factor that influences the Cherenkov photon density. At high zenith angles, that is, at high angles with respect to the vertical to the ground, the shower passes through a larger atmosphere layer. Since the distance to the point of maximum development of the shower increases, the photon density decreases with increasing fluctuations. This effect is important, especially for low-energy primary gammas.

\begin{figure}[b!]
\centering
\includegraphics[width=0.65\linewidth]{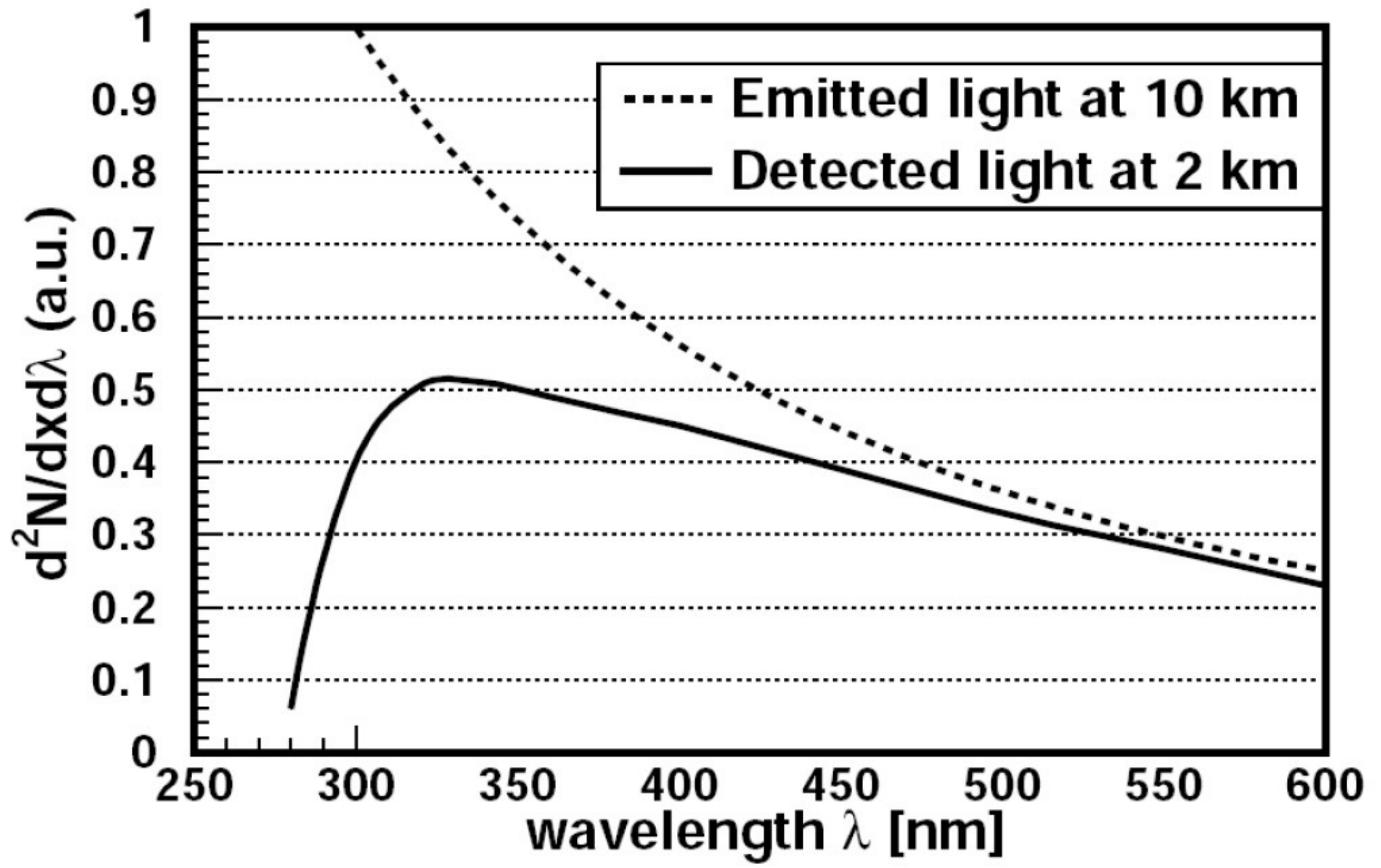}
\caption[Effect of the absorption of Cherenkov light in the atmosphere.]{Effect of the absorption of Cherenkov light in the atmosphere.}
\label{img:CHERspectrum}
\end{figure}

Cherenkov radiation emitted by shower particles has a typical spectrum ranging from 300 to 600\,nm. The emitted spectrum, shown with the dotted line in Figure~\ref{img:CHERspectrum}, is attenuated in the atmosphere. 

The total luminosity observed by a Cherenkov telescope is mostly affected by the following contributions:
\begin{itemize}
    \item  the Night Sky Background (NSB);
    \item the Moon;
    \item the Earth magnetic field that tends to deviate the trajectories of charged particles from atmospheric showers, especially those with lower energies.
\end{itemize}

\subsubsection{The imaging technique}
\begin{figure}[htbp]
\centering
\includegraphics[width=0.6\linewidth]{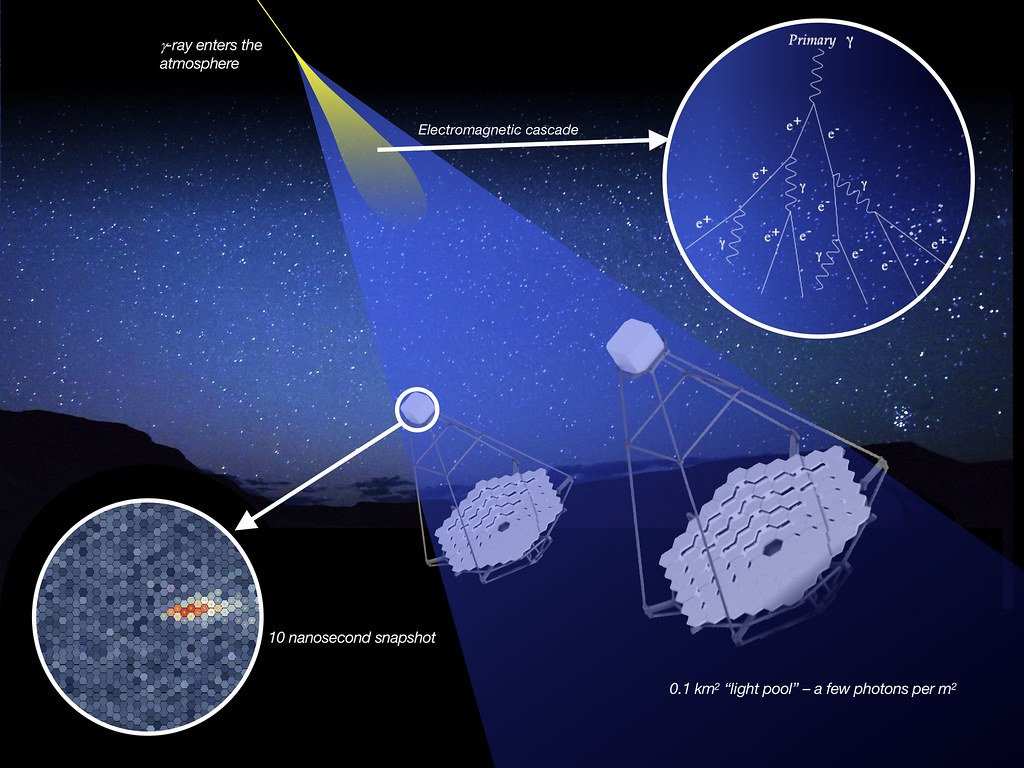}
\caption[Process of image formation for an IACT.]{Schematics of IACT detection of atmospheric showers, from\cite{cta-exp}. 
}
\label{img:iact_principle}
\end{figure}
Identification of the differences between electromagnetic and hadronic cascades is one of the primary targets of $\gamma$-ray ground-based telescopes. 
The ratio of $\gamma$--rays to charged cosmic rays is, in fact, really small ($\sim 1\times 10^{-4}$), and a very powerful technique is needed to separate the gamma events from the dominating hadrons.
A successful method is the imaging technique. 
The technique is based on the study of the images produced by Cherenkov photons produced by EAS when focused on a plane.

The entire image formation process is shown in Figure~\ref{img:iact_principle}: Cherenkov photons are distributed on a surface of \emph{elliptical shape}, whose extremities represent the head and tail of the shower, while the inner pixels are the core of the shower. The basic idea of the imaging technique is to use the shape and orientation of the images to extract physical information about the primary particle.

The elliptical image can be parameterized by a set of parameters, the {\it image parameters}, which are the indispensable tool of the imaging technique for the characterization of the primary particle. The primary scope of an IACT is that of collecting shower images, focusing the Cherenkov light emitted by atmospheric shower particles during their passage through the atmosphere. The necessary hardware elements for this kind of instrument are: a rotating structure, essential for the tracking of gamma rays emitter candidates;  a reflective surface, which focuses the light; a camera, where the light forming the image is focused and the signal registered; a trigger, producing a first rough background rejection; and a data acquisition system, which provides the signal storing for the subsequent analysis. 

\begin{figure}[h!]
\centering
\includegraphics[width=0.7\linewidth]{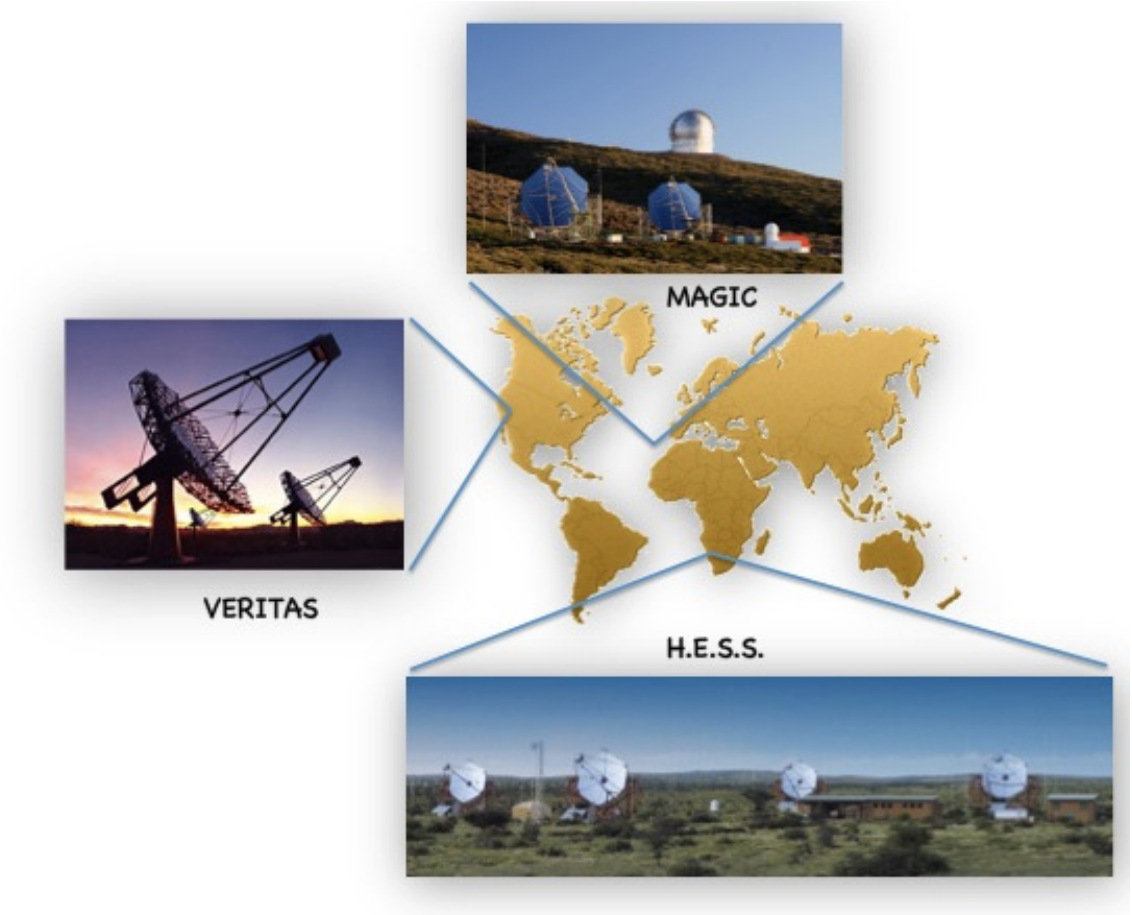}\caption{Location of the
three major IACTs currently in operation: H.E.S.S., MAGIC and VERITAS.}\label{fig:iact_world}
\end{figure}

There are four operating IACTs, namely, H.E.S.S. \cite{hess-exp} in Namibia, MAGIC \cite{magic-exp} and FACT \cite{fact-exp} in La Palma, Canary Island, and VERITAS in Arizona \cite{veritas-exp}. The locations of H.E.S.S., MAGIC, and VERITAS are illustrated in Figure~\ref{fig:iact_world}.

\begin{figure}
    \centering
    \includegraphics[width=0.6\linewidth]{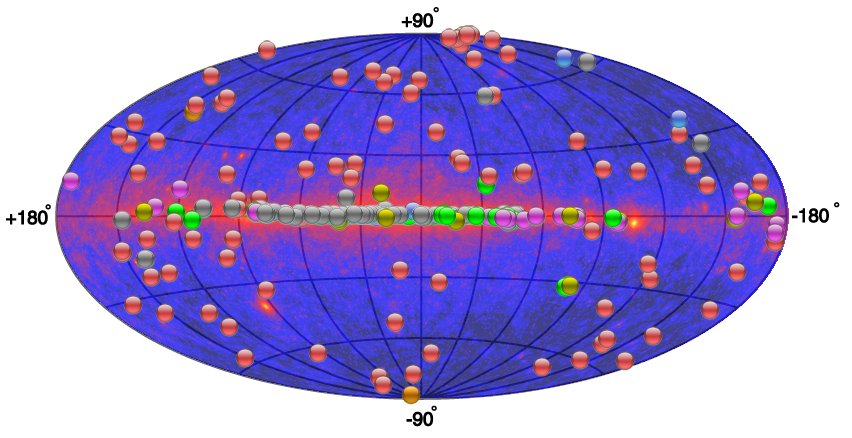}
    \includegraphics[width=0.14\linewidth]{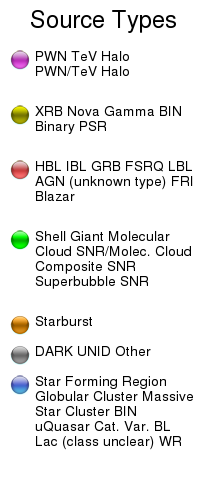}
    \caption{Map of the location of the TeV-detected sources superimposed to the {\textit Fermi}-LAT full sky map. The source type is reported in the legend. From TeVCat online catalog \citep{tevcat}.}
    \label{fig:VHEsky}
\end{figure}

Future instruments, such as MACE \cite{mace-exp}, ASTRI \cite{astri-exp}, and CTA \cite{cta-exp}, are currently in the construction phase. Figure~\ref{fig:VHEsky} illustrates the map of the TeV sky detected in the last 30 years of IACT observations. It counts $\sim$ 250 sources, approximately 1/3 extragalactic and 2/3 galactic. The future CTA is expected to bring a revolution in the field, bringing the number of sources from fewer than 300 to several thousands. This will be accomplished by the construction of two sites, one in the north and one in the south, and the deployment of a large number of telescopes of 3 different sizes.

\newpage

\subsection{Direct detection of EAS: high-altitude particle detectors}

Direct detection of air shower particles offers a method of VHE $\gamma$-ray detection with a duty cycle close to 100\,\% and a very wide field of view. The technique was adapted from EAS cosmic-ray detectors by using denser arrays of particle detectors located at higher altitudes. The energy threshold reached with this technique is around 1 TeV. 

\begin{figure}
    \centering
    \includegraphics[width=0.3\linewidth]{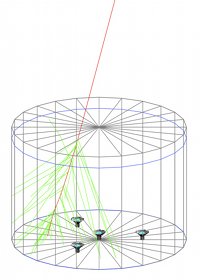}
    \includegraphics[width=0.5\linewidth]{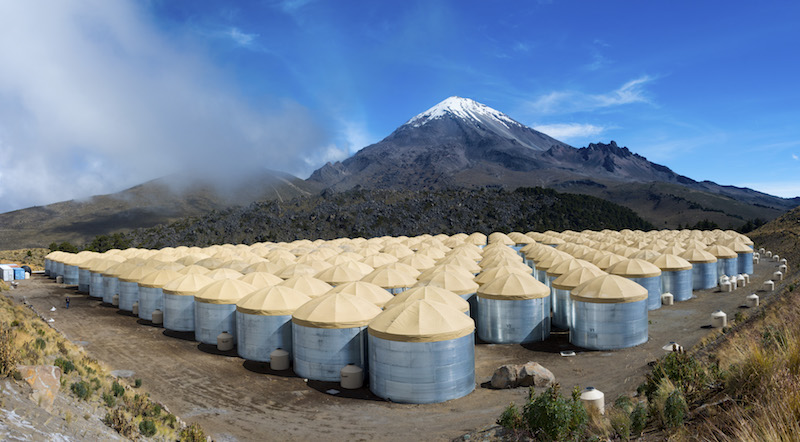}
    \caption{Water Cherenkov technique for EAS detection. Left: Simulation of a charged particle passing through a tank (red line) and emitting Cherenkov light (green lines). Also shown is the layout of the light sensors of the HAWC experiment, from the HAWC public website. Right: The HAWC Observatory (J. Goodman, Nov. 2016).}
    \label{fig:hawc}
\end{figure}

The three main techniques adopted to measure and characterize charged particles of EAS initiated by gamma rays are: \begin{itemize} 
\item Cherenkov light detection. This can be achieved with detector units spread over a large surface, for example, the HAWC Observatory \cite{hawc-exp}, which has been operating since 2016, or with segmented ponds, for example, the Milagro Observatory, which operated from 2000 to 2008 in New Mexico. HAWC is located in Sierra Negra, Mexico, at 4100 m above sea level and is composed of 300 water Cherenkov detectors (Figure~\ref{fig:hawc}). In HAWC, purified water tanks a few meters deep and instrumented with few light sensors are adopted.  
\item Scintillation light detection. An example is the Tibet Air Shower (AS)-gamma Experiment \citep{tibet-as-array}, operating in Tibet.
\item Resistive plate chamber detection. An example is the ARGO experiment \cite{argo-exp}, in operation from 2001 to 2013, located on the same high-altitude plateau in Tibet. 
\end{itemize}

The main challenges of the direct detection approach are the rejection of the cosmic-ray background, based on the muon content of showers and/or the distribution of shower particles on the ground, and directional and energy reconstruction. The large fluctuations present in the particle number at ground level make primary energy determination extremely difficult for extended atmospheric shower detectors.

To reduce the uncertainties related to the large distance from the detector to the shower maximum, the next generation of this kind of detector will be built at altitudes above 4000 meters and will complement, with its wide FoV and large duty cycle, the next generation of IACTs. Preliminary results presented by the LHAASO collaboration suggest that the rejection capability of the cosmic-ray background might be greatly improved with muon detectors placed in the telescope area.

\begin{figure}
    \centering
    \includegraphics[width=0.7\linewidth]{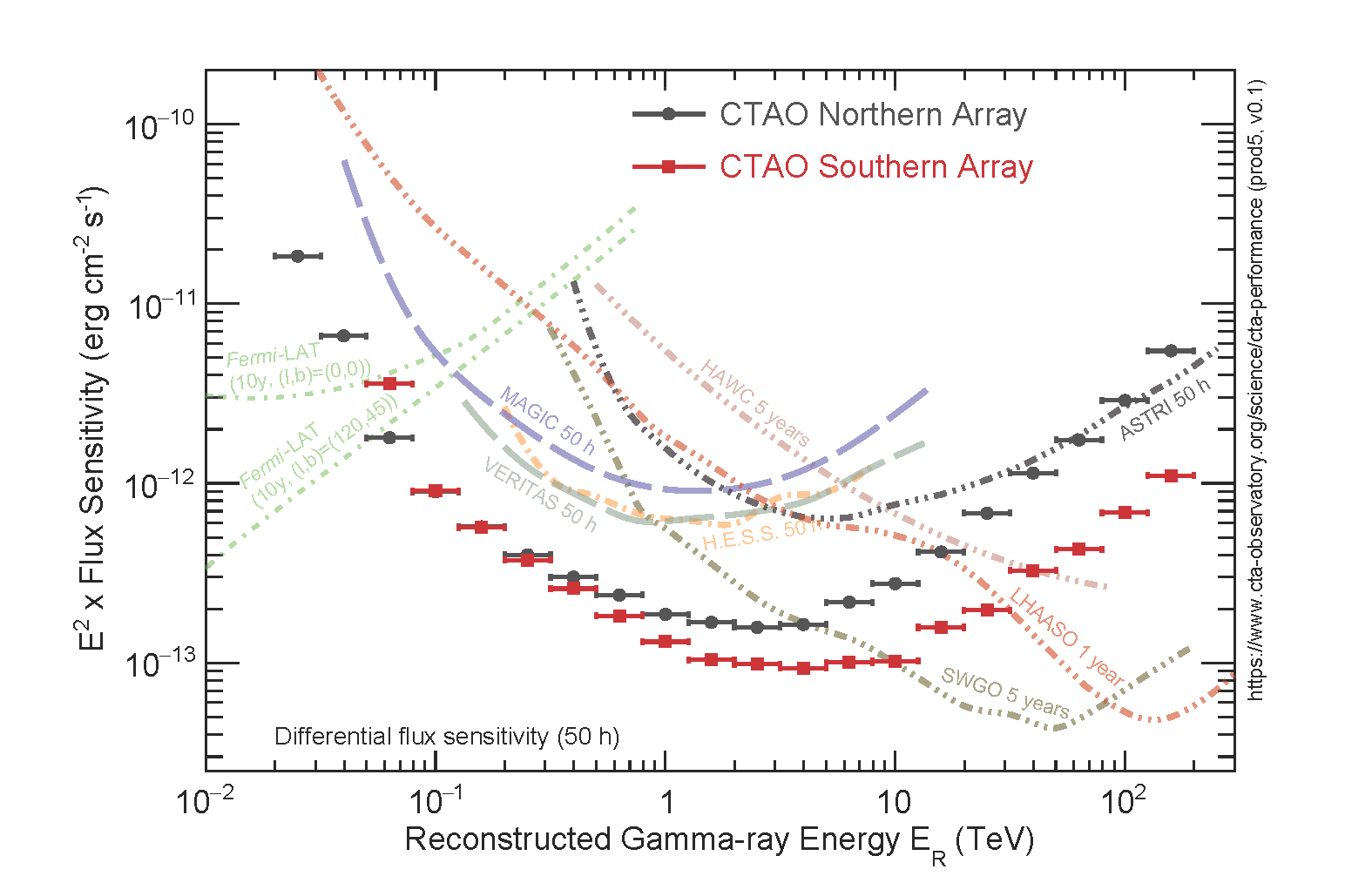}
    \caption{Differential sensitivity of current and future generation of IACT and EAS detectors. CTA collaboration (https://www.ctaobservatory.org/science/ctao-performance/)}
    \label{fig:diffsens_VHE}
\end{figure}

LHAASO \cite{lhaaso-exp} is currently in the construction phase, but started scientific operations in 2019. LHAASO is a gigantic array of detectors located at 4410\,m altitude in China. Thanks to the combination of different detector types, namely, 5195 scintillation units, 3000 water Cherenkov units, 12 wide field Cherenkov telescopes, and a dedicated array of 1171 muon detectors for background rejection, LAHAASO is expected to reach unprecedented sensitivity over a broad energy range; see Figure~\ref{fig:diffsens_VHE}. First, exciting results were obtained with half of the array in 2020, reporting the detection of a dozen PeVatrons in our galaxy and inaugurating the Ultra-High-Energy Gamma Astronomy Era \cite{2021Natur.594...33C}. 

LHAASO, located in China, is best suited for observation of the northern sky. For the southern sky, a new collaboration, named SWGO \cite{swgo-exp}, was recently formed. The SWGO Collaboration plans to build an array of high-altitude EAS detectors for gamma-ray astronomy in South America and is currently in the R\&D phase.

\newpage
\section{Gamma-ray propagation from extragalactic sources: the standard scenario}
Once escaped from the source, gamma rays propagate in the galactic/extragalactic space. Depending mainly on three factors,  i.e., their energy, distance, and the properties of the medium itself, their travel might be ``disturbed''. 
This has a direct effect on the spectra and/or lightcurves detected by gamma-ray telescopes. Interestingly, this gives an opportunity to probe the disturber field/process, namely, the extragalactic background light, the intergalactic magnetic field, gravitational lensing galaxies, or quantum gravity modification effects. 

In this section, we present the main effects that affect gamma-ray propagation from extragalactic emitters. We limit the discussion to the effects not deviating from the standard model (SM), and leave to Section~\ref{Sec:GR prop:alternatives} the description of additional effects related to non-SM theories, such as quantum gravity and axions.
Propagation effects affecting galactic emitters are discussed in the next section. 

\subsection{Extragalactic background light}
Gamma rays are an ideal messenger of HE astrophysical phenomena, as they point to their emitter and are relatively easy to detect (from space or the ground). The caveat is that they might be absorbed during their travel towards Earth from their point of origin. Let us assume that we have a HE gamma ray traveling from a distant source, e.g., a blazar (subclass of AGNs whose relativistic jets point towards Earth),  to us. What kind of interaction can it experience? The dominant process that we have that may affect gamma-ray propagation is electron-positron pair production.
The net effect is that the flux observed from Earth is suppressed by a factor that depends on the opacity ($\tau$). The opacity itself depends on the emitter's distance and the energy of the VHE photon. If a source is farther away, the VHE photon will have more chances to interact simply because, on its way, it has more ``meetings'' with other photons, and thus has more chances to interact. Similarly, the higher the energy, the higher the probability of the interaction. In the following, $\epsilon$ denotes the energy of the background photon needed for the electron-positron pair production
\begin{equation}\label{eq:observed_spectrum}
\left(\frac{d N}{d E}\right)_{o b s}=\left(\frac{d N}{d E}\right)_{e m} \cdot e^{-\tau\left(z_{e}, E\right)}\,.
\end{equation}
Let us take a look at the numbers. Imagine that we have a photon of 1 TeV coming from a blazar. It encounters a background photon, and they create an electron-positron pair. This process is allowed kinematically above the energy reaction threshold. At a gamma-ray energy roughly twice the energy threshold, the cross section has its maximum \cite{2013MNRAS.432.3245D}. For a head-on collision, the minimal energy of the background photon for electron-positron pair creation with 1\,TeV gamma ray is 0.5\,eV:

\begin{equation}\label{eq:epsilon_max}
\epsilon_{th} \sim \frac{2\left(m_{e} c^{2}\right)^{2}}{E_{\gamma}} \sim 0.5\left(\frac{1 \mathrm{TeV}}{E_{\gamma}}\right) \mathrm{eV}\,.
\end{equation}
In other words, if we look at the wavelength, it means that 1\,TeV photon interacts with a photon of around 1 $\mu m$ 
\begin{equation} \label{eq:lambda_max}
\lambda_{\max } \sim 1.24\left(E_{\gamma}[\mathrm{TeV}] \mu m\right)\,.
\end{equation}

What kind of background are we talking about here? 
\begin{figure}
    \centering
    \includegraphics[width=0.9\textwidth]{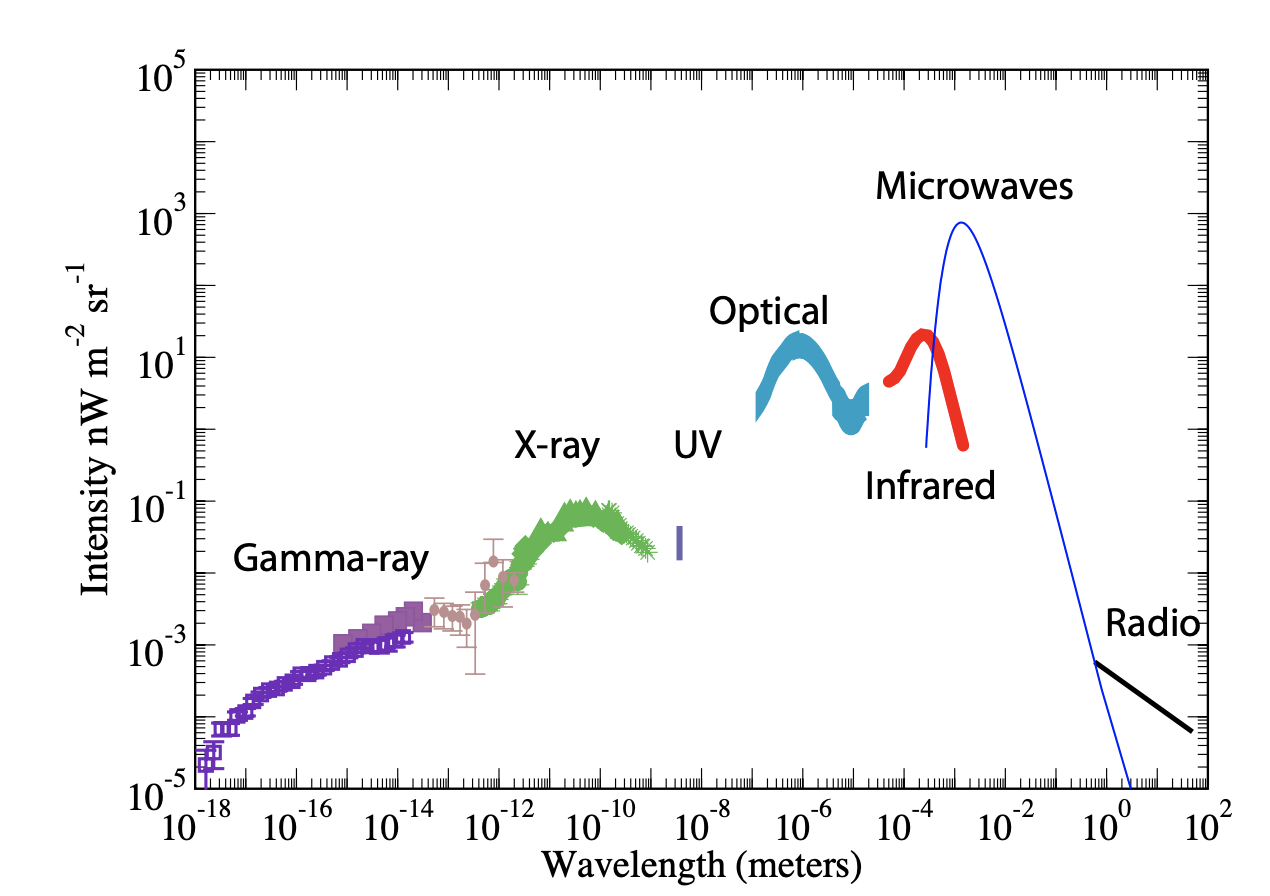}
    \caption{Intensity of the extragalactic background as a function of the wavelength. \cite{Cooray2016ExtragalacticBL}}
    \label{fig:Backgrounds_in_the_Universe}
\end{figure}
If we look at Figure~\ref{fig:Backgrounds_in_the_Universe} we see that we have the Cosmic Microwave Background (CMB), which is very bright and then we have all the other backgrounds that are filling the universe. The essential backgrounds for our 1\,TeV photon are optical and infrared, not CMB, because the energy is not high enough. CMB is a perfect Planckian, and it is a picture of a specific phenomenon at a specific time. What happened to the CMB after its appearance? Nothing, it simply diluted. It is not the same for the other backgrounds. If we look at any other background, e.g., optical produced mainly by the stars, we see that this light is not an image of a phenomenon in a specific moment of the Universe, like CMB. Namely, the stars were exploding and then dying, exploding again and dying, and now we observe the third population of stars. 
Therefore, by looking at the optical-IR background (the so-called extragalactic background light, EBL), we are looking at all the history of the star production and dust reprocessing of this light (the main reason for IR light). This is the first reason why it is so challenging to study the EBL. It contains a lot of information about the cosmic evolution of sources, and we do not know much about that. The EBL spectrum captures the redshifted energy released from all stars and galaxies throughout cosmic history, including first stellar objects, primordial black holes, and proto-galaxies. So looking at the EBL means looking at the photons collected at the first stars and the last stars, all together. 

EBL is the second brightest diffuse background radiation permeating the Universe. It is 30-40 times less bright than the CMB. It is poorly known experimentally, especially in the IR band. On the contrary, CMB has been measured with a precision better than 1\%. The challenge lies in removing the foregrounds as zodiac light. That is, we need to subtract a background that is dominant with respect to the EBL. Therefore, the EBL that we take in our photon-photon interaction is an entire research field on its own; we need to remember that. 
The characteristics of the EBL are two peaks with comparable intensity. The first peak is in the optical band (1$\mu$m), and the second peak is in the IR band (100 $\mu$m). Thermonuclear burning in stars (90\%) and AGN accretion (10\%) are the main responsible for the optical part. The optical light is reprocessed over cosmic time by dust and emitted as IR. 

How do we measure EBL? We use direct and indirect detection \citep[for a detailed review, see][]{2001ARA&A..39..249H}. Direct measurements obtained exciting results in the $1-5 \mu m$ region. That is, a team of researchers found a hint of a flux peak there \cite{2005ApJ...626...31M}. Primaeval (Population III) stars, the first stars in the Universe, have been considered as a possible origin. The peak turned out to be due to strong foreground emission that has not been properly subtracted. 
In the band where we have the valley of the EBL ($5-100 \mu m$) we have a peak in the foreground emission (nearly independent of the ecliptic coordinates), so it becomes even more challenging to measure EBL directly. 
Above 100\,$\mu$m we have more reliable measurements thanks to a better known background (at least until we have the CMB popping out and we again have domination of a background). 
One way to solve or circumvent this problem is to study the background fluctuations of the emission we receive with respect to the average emission. Here, we rely on the fact that the Zodiacal light is uniform on scales lower than a degree. In this case, the background subtraction might be non-trivial, and the detector might contribute (instrumental point spread function).

We saw that direct EBL measurement is notoriously difficult, but what about indirect measurements? Galaxy counts provided solid lower limits in the optical bound. The main idea is the following: the light emitted by the galaxies that we see is the minimum optical light diffuse in the Universe. In the EBL measurement, we cannot forget that the EBL that we see now is the superposition of the whole history of the EBL altogether. So, if we want to go back in time, we face the challenge of modeling how our photon emitted at redshift $z$ evolved during time. It did not interact with what we now see of the EBL. It interacted with the previous EBL. The reason for this is simple: stars (and dust) evolved; therefore, it is hard to model the interaction, since the EBL changes over time.
\begin{figure} [h!]
    \centering
    \includegraphics[width=0.9\textwidth]{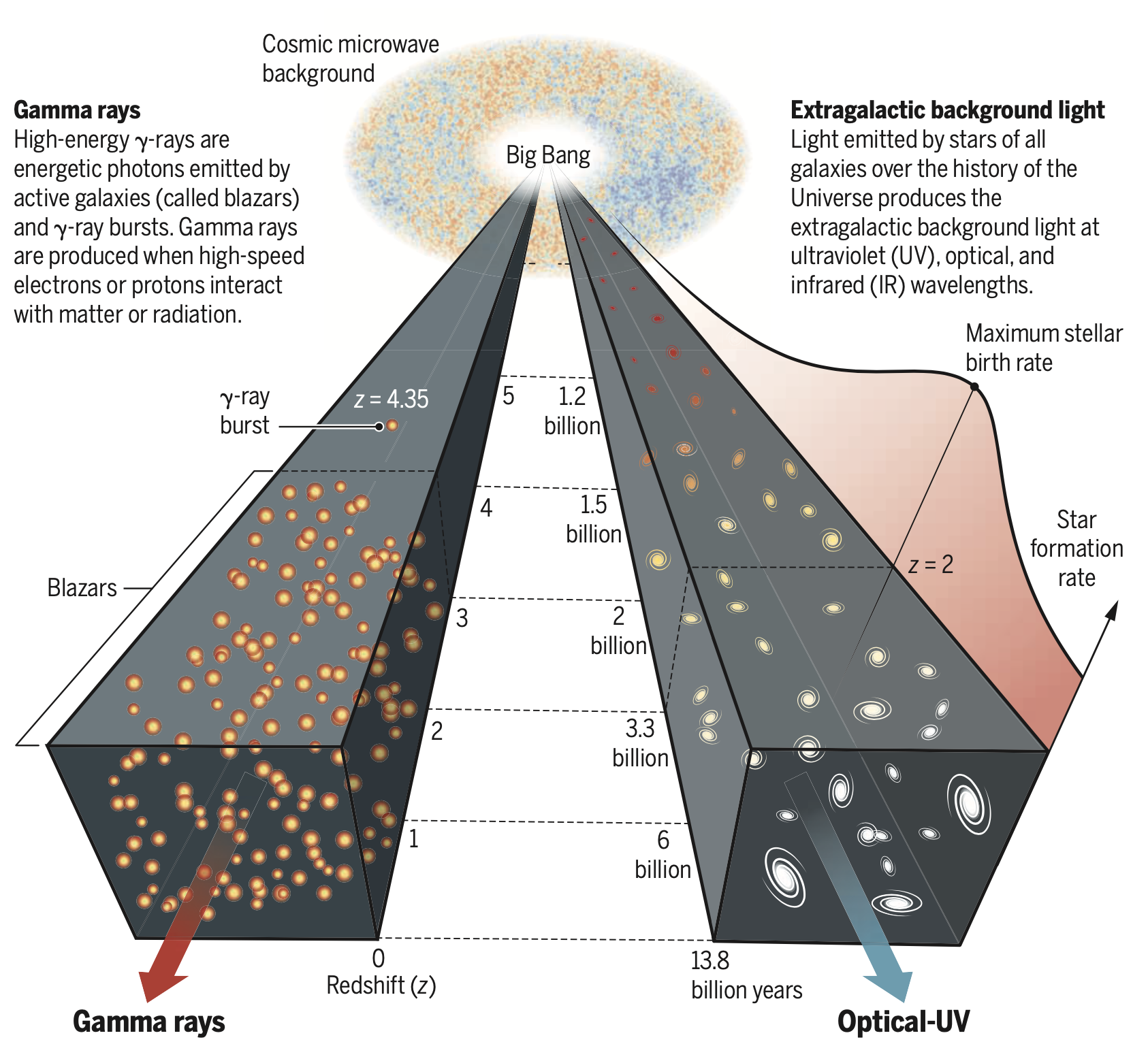}
    \caption{EBL evolution, represented on the right, evolves over time following star and galaxy evolution. Gamma rays from extragalactic emitters (e.g., blazars or GRBs) might interact during their travel with the EBL \cite{Fermi-LAT:2018lqt}.}
    \label{fig:Redshift_evolution}
\end{figure}

\begin{figure} [h!]
    \centering
    \includegraphics[width=0.9\textwidth]{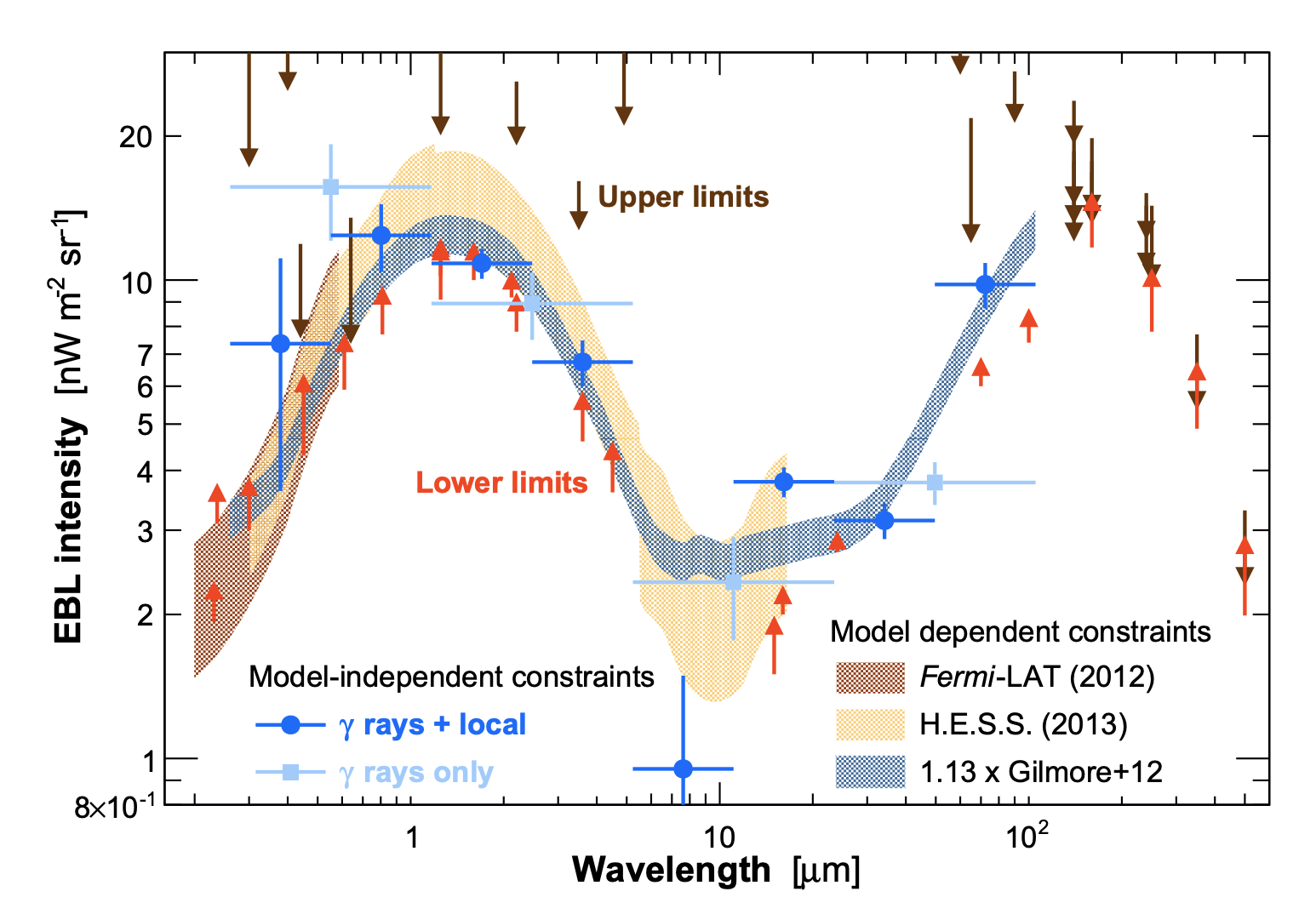}
    \caption{EBL intensity at $z = 1$.  \cite{2015ApJ...812...60B}}
    \label{fig:EBL_Biteau}
\end{figure}

In Figure~\ref{fig:Redshift_evolution} we can see that the peak of star formation is around $z=2$. Now, we live in a Universe that is much less active in stellar formation, and this means that the EBL changed a lot. EBL did not change just because the stellar count changed, but also because the dust, which is processing and reemitting this light, has changed. Ideally, modeling the EBL means: accounting for the observed data at different redshifts, matching the current (z = 0) EBL intensity, and following the physical prescriptions, e.g., on the history of star formation rate in the Universe. In reality, there are different models.
\textbf{Empirical models} are data-driven. Such models are, for example, Franceschini \cite{Franceschini:2017iwq}, Dominguez \cite{Dominguez:2010bv}. In the Franceschini model, the authors tried to model the current EBL and then proposed the evolution with the time needed to compute the absorption factor.  \textbf{Physical models} have a priori assumptions based on physics. Such models are, for example, Primack \cite{Primack:2011ny} and Gilmore \cite{Gilmore:2011ks}. There are \textbf{other models} as well built on the parameterization of the star formation rate. Regardless of the method, all EBL models are in good agreement and have been tested on VHE gamma rays by IACTs \cite{HESS:2017vis,MAGIC:2019ozu,VERITAS:2019ria}. In Figure~\ref{fig:EBL_Biteau} EBL intensity at redshift $z=1$ is depicted.

We can now proceed to compute the cosmic opacity of the gamma rays. 
For close-by sources, this EBL absorption effect occurs only at high energies, around 40\,TeV. If we go a bit further, for example, at $z=0.1$, the absorption is more effective earlier, around 1\,TeV. Moreover, after 1\,TeV, this absorption ``explode'', and as a consequence, our photons are strongly suppressed. This means that our flux is increasingly suppressed at higher energies. For higher energies, at this distance, the survival probability is almost 0. What happens if we go to redshift $z=1$, which is the current limit of the IACT observations? IACT threshold is around 100\,GeV, and at such distances $\tau$ reaches unity at 100\,GeV. As a consequence, the flux is strongly suppressed after 100\,GeV. Because of this, if we want to study the distant Universe, we must focus on the HE band (E$<$100\,GeV); at VHE, the suppression is too strong. 
The energy of a gamma-ray photon at which $\tau$ reaches unity at a particular redshift is called the gamma-ray horizon.

In our experiments, we measure the spectrum, assume an EBL model, and know the redshift so that we can reconstruct the intrinsic spectrum. This process is called deabsorption. In this way, we can study the property of the spectral energy distribution. Below a few tens of GeV, we do not have the EBL absorption process as we are below the threshold energy for pair creation. 
To estimate the intrinsic spectrum, we need the EBL model, the distance of the source, and the measured spectrum. However, we can also reverse the reasoning. We can have an observed spectrum, assume an intrinsic spectrum (usually a power law) and the redshift, and using that information, we can estimate the EBL. This is one of the methods developed with IACT data to constrain the EBL \cite{HESS:2017vis,MAGIC:2019ozu,VERITAS:2019ria}. Many authors have been able to set the EBL limits in this way. The good news is that these results are close to the lower limits set by the galaxy count. So what does it mean? It means that the galaxies that we see are probably all the galaxies of the Universe; we do not lack much light. 

Extreme blazars can be used as powerful probes for the EBL. They are sources with a bump in the spectral energy distribution that extends beyond 1,TeV. This means that we have a lot of TeV photons. If they are close enough (moderate redshift $z < 0.2$), they are ideal for probing IR EBL, for which we need nearby sources and a lot of TeV photons. The caveat is that they are faint sources. 

What happens if, in the process of deabsorption, we correct our spectrum too much? If we take an EBL that is too luminous, our spectrum will show an upturn at the highest energies, and it will become unphysical. This approach is used to set limits on the blazar distance (which is often unknown). 

In summary, EBL plays a significant role in the propagation of extragalactic photons at tens of GeV (distant sources, $z > 1$) and TeV sources (all distances). It is poorly known because of foregrounds. EBL and its evolution over time can be modeled and is related to cosmology. GeV-TeV data can be used to test EBL models, constrain intrinsic emission, or set the source distance. 

\subsection{Intergalactic magnetic field}
What if we add a magnetic field to the game? The magnetic field will affect electrons and positrons created in photon-photon interaction and electron-positron pair production. It deviates pairs and opens them. These electron-positron pairs then might up-scatter CMB photons via the inverse Compton process. These secondary photons will either produce new electron-positron pairs (if they have sufficient energy to reach the energy reaction threshold) or propagate undisturbed through the Universe on their way to Earth. The thing is that the intragalactic magnetic field is mainly unknown (by order of magnitude). So, depending on the strength, we can imagine that our deviation will be more substantial or smaller. So why not use those observations to probe the magnetic field? There are three main gamma-ray observables relevant to intergalactic magnetic field studies: spectral effects, angular distribution, and time delays (the secondary photons will be delayed compared to primary gamma rays). More information on this topic can be found in a review by Batista et al. \cite{Batista:2021rgm} and references therein. 

\subsection{Gravitational lensing}
Another effect that might affect gamma-ray propagation by cosmological distances is gravitational lensing. This phenomenon characterizes the propagation of photons that encounter a large mass (e.g., a galaxy) while traveling from the source to the observer, with the result that the trajectory of the photons is deviated. Interestingly, depending on the geometry of the system, multiple trajectories are possible, and a delay in the arrival time due to a different effective path length traveled by the photons can become detectable. 
Only a handful of gamma-ray sources, all blazars, have shown this effect \cite[e.g.,][]{2014ApJ...782L..14C}. One of these sources was detected by an IACT, at VHE gamma rays \cite{2016A&A...595A..98A}.

\section{Gamma-ray propagation from galactic sources: the standard scenario }

In the previous chapter, we saw how the EBL influenced the propagation of VHE gamma rays. The farther the distance, the higher the attenuation. However, what is with the attenuation in our galaxy? Now, the VHE gamma ray absorption targets are radiation fields in the Milky Way. When dealing with the EBL, we did not care about the direction of our source; i.e., the optical depth was calculated for all the sources of the same redshift in the same way. The situation of galactic absorption is different, and the density, energy, and angular distribution must be known to properly model galactic absorption \cite{Moskalenko:2005ng, Vernetto:2016alq}. 

The background radiation field in the Milky Way has four components. These are CMB and EBL (extragalactic components) and radiation emitted by stars and dust. The most important source of target photons (with a photon number density of $410\,\mathrm{cm}^{-3}$) is the CMB. It is mainly responsible for the attenuation of gamma rays from a few hundred TeV to a few hundred PeV. At energies below 300 TeV, the most important targets for gamma-ray attenuation are infrared photons with wavelengths in the range $\lambda = 50 - 500 \mu\mathrm{m}$. These low-energy photons are the result of dust emission, heated by stellar light. One can legitimately ask: What about the starlight photons; How do they contribute to the total VHE gamma-ray galactic attenuation? Starlight photons are most effective in absorbing gamma rays with energy $\sim 1$ TeV. Because starlight photons have a lower number density (0.42 photons/$\mathrm{cm}^{-3}$) in comparison with photons emitted by dust (24.9 $\mathrm{cm}^{-3}$ ), their contribution to the total gamma-ray galactic attenuation can be considered negligible (smaller than 2\% for gamma rays created in any point of the Milky Way).
According to some studies (e.g. \cite{Vernetto:2016alq}) the absorption probability due to Galactic infrared radiation is maximum for gamma rays with energies $\sim 1$ TeV. If the line of sight is passing close to the Galactic center, this absorption probability can be 0.45. One of the main problems for modeling the absorption probability is getting an analytic description of the interstellar dust. 

\begin{figure} [h!]
    \centering
    \includegraphics[width=0.48\textwidth]{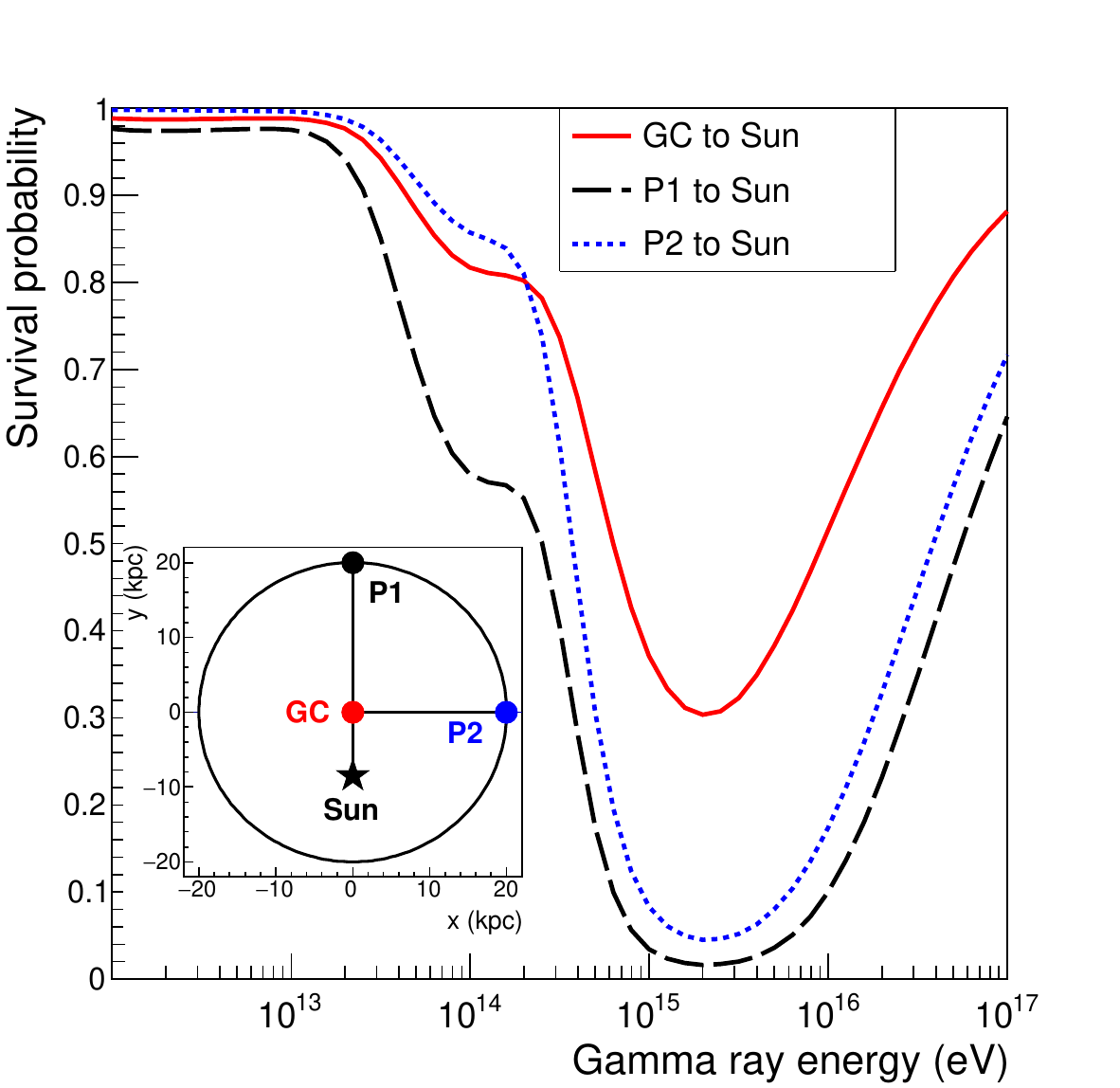}
    \includegraphics[width=0.48\textwidth]{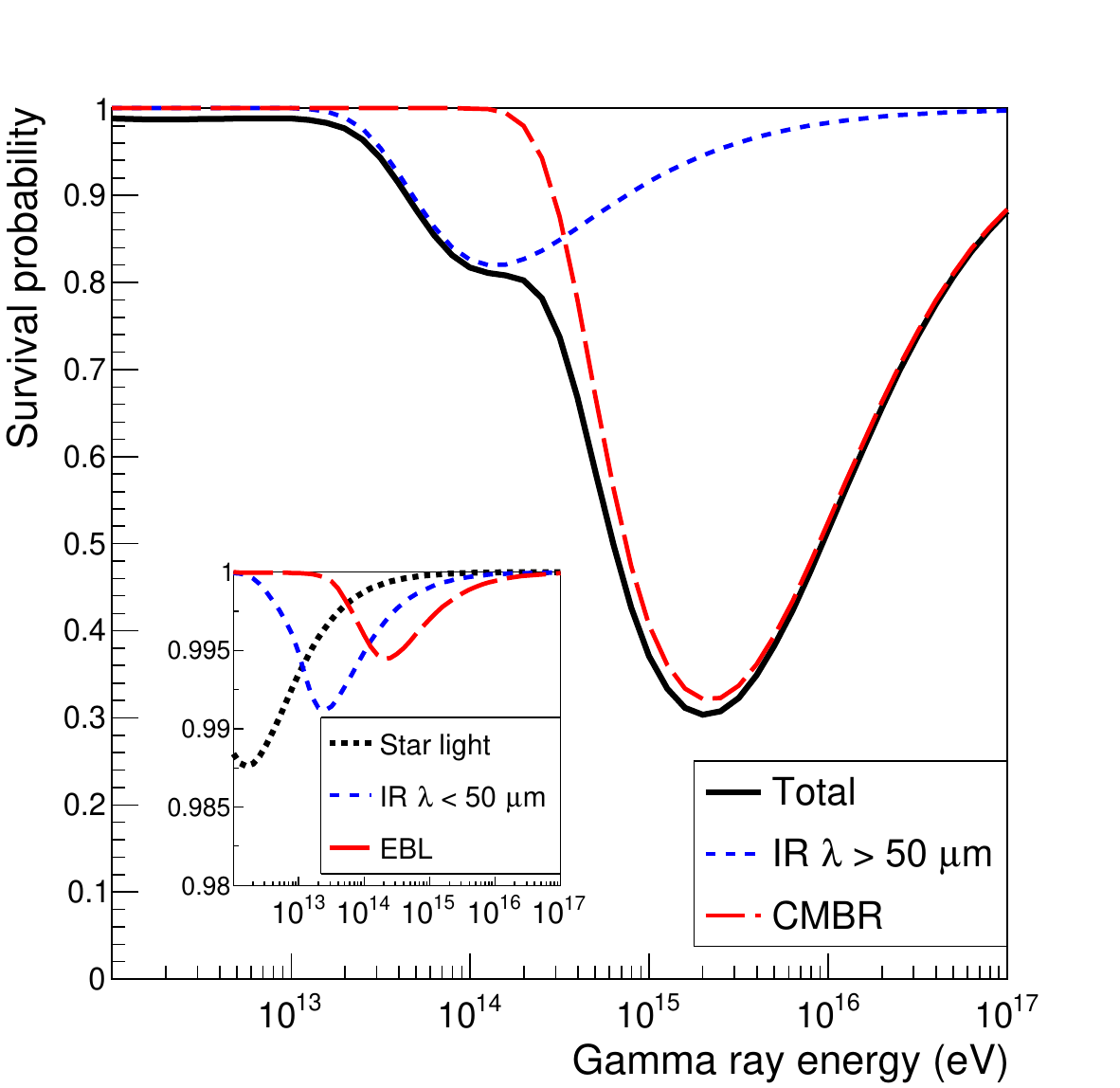}
    \caption{\textit{Left:}Survival probability of gamma rays for a trajectory from the Galactic center to the Sun \textit{Right:} Survival probability of gamma rays for three different trajectories in the Galactic plane. The inserted plot shows the position of the sources. \cite{Vernetto:2016alq}.}
    \label{fig:Verneto_Lipari}
\end{figure}

In the left part of Figure~\ref{fig:Verneto_Lipari} we can see the survival probability of gamma rays coming from the source situated in the Galactic center to the Sun as a function of the gamma ray energy. We can see that the main contributors to the absorption are the CMB and the thermal emission from the dust. In the right part of Figure~\ref{fig:Verneto_Lipari} the survival probability for three different source positions as a function of gamma-ray energy is depicted. In the inserted plot, we can see the exact distance of those sources from the Galactic center, in kpc. The absorption is maximal when the gamma rays need to cross the Galactic center. Looking at the optical depth of the minimum survival probability at 150 TeV, we can see that infrared absorption is maximal in this scenario. 

\section{Gamma-ray propagation: alternative scenarios}
\label{Sec:GR prop:alternatives}
We saw that there are many ways to study quantum gravity with gamma rays. First, we can do this by looking at the time of arrival of photons. To do so, we need light curves. If we want to see a difference in the arrival times, we need to build light curves as finely as possible. Another way to do quantum gravity or search for Lorentz invariance violation (LIV) with gamma rays is a study of spectral energy distribution and, in particular, the deviation from the standard scenario expected for photon propagation. In this case, we are trying to see the shape of the spectrum that enables us to disentangle the different models.

\subsection{Lorentz invariance violation}
We start with the modified photon dispersion relation when we want to test the LIV with some gamma ray data. It is a simple way of parameterizing the ``out-of-ordinary'' behavior.

\begin{equation} \label{eq:MPDR}
E^{2} \simeq p^{2} \times\left[1-\sum_{n=1}^{\infty} S\left(\frac{E}{E_{\mathrm{QG}, \mathrm{n}}}\right)^{n}\right]\,.
\end{equation}

$E_{\mathrm{QG}, \mathrm{n}}$ denotes the energy scale at which QG effects are expected to be significant, $E$ is the energy of a VHE gamma ray. $S$ can have values of +1 and -1. We usually test for the linear contribution $(n=1)$ and quadratic $(n=2)$. Non-integer values are also possible, and in this case, we are talking about fractional models \cite{Calcagni:2016zqv}.  
Everything starts with Eq.\,\eqref{eq:MPDR}, but what are the consequences of modifying the dispersion relation? We have many lines of research on QG effects. Some of them are Time-of-flight studies, Space-time fuzziness, Modification of gamma-gamma cross section (with EBL or in atmospheric shower development, i.e., Bethe-Heitler process), Photon decay and Photon splitting (only possible in the superluminal scenario), etc.  

\subsubsection{Modification on the propagation: optical depth}

\begin{figure} [h!]
    \centering
    \includegraphics[width=0.9\textwidth]{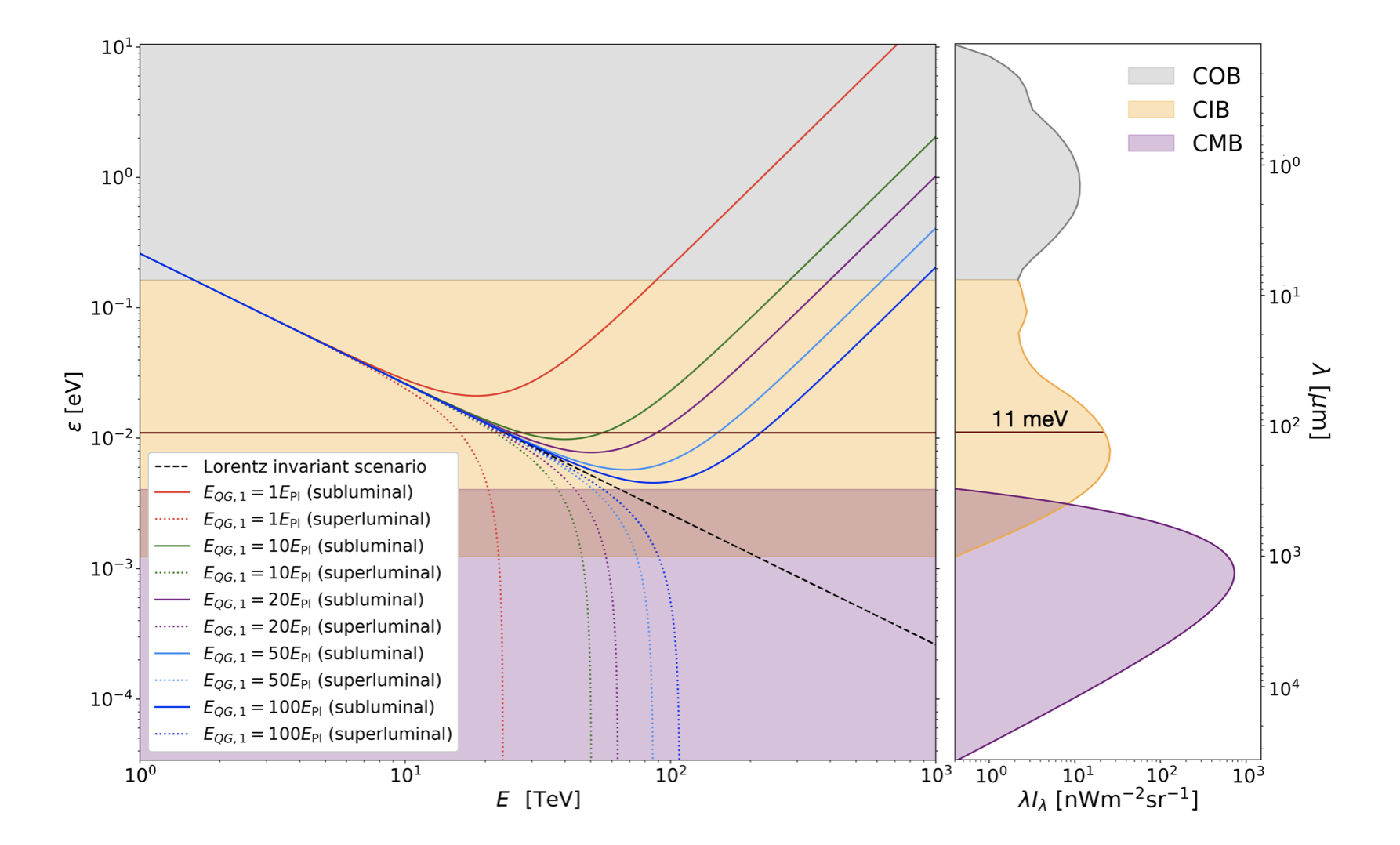}
    \caption{\textit{Left:} Energy of the background photons at the reaction energy threshold for the electron positron pair production a function of a VHE gamma-ray energy. The black dashed line represents the Lorentz invariant scenario, solid lines represent LIV subluminal and doted LIV superluminal scenarios for five different values of QG scale. EBL photon of 11~meV is depicted with the solid maroon line so that two reaction energy thresholds in the LIV subluminal scenario, and consequently energy domain of VHE gamma-rays capable of interaction, are easily visible \textit{Right:} Spectral energy distributions of the CMB and EBL. From \cite{Terzic:2021rlx}  }
    \label{fig:Threshold_SED}
\end{figure}

If the dispersion relation is modified, the reaction energy threshold also changes. In the LIV superluminal scenario, the reaction energy threshold is lower (dotted lines in Figure~\ref{fig:Threshold_SED}); thus, it is easier to reach the reaction energy threshold than in the Lorentz-invariant scenario. Moreover, VHE gamma rays (e.g., E = 50 TeV), which in the Lorentz invariant scenario would not interact with CMB photons, are now interacting with it for sufficiently low values of the QG energy scale (e.g., green dotted line in Figure~\ref{fig:Threshold_SED}). For those reasons, the attenuation will be more pronounced in the LIV superluminal scenario than in the Lorentz invariant.  

\begin{figure} [h!]
    \centering
    \includegraphics[width=0.5\textwidth]{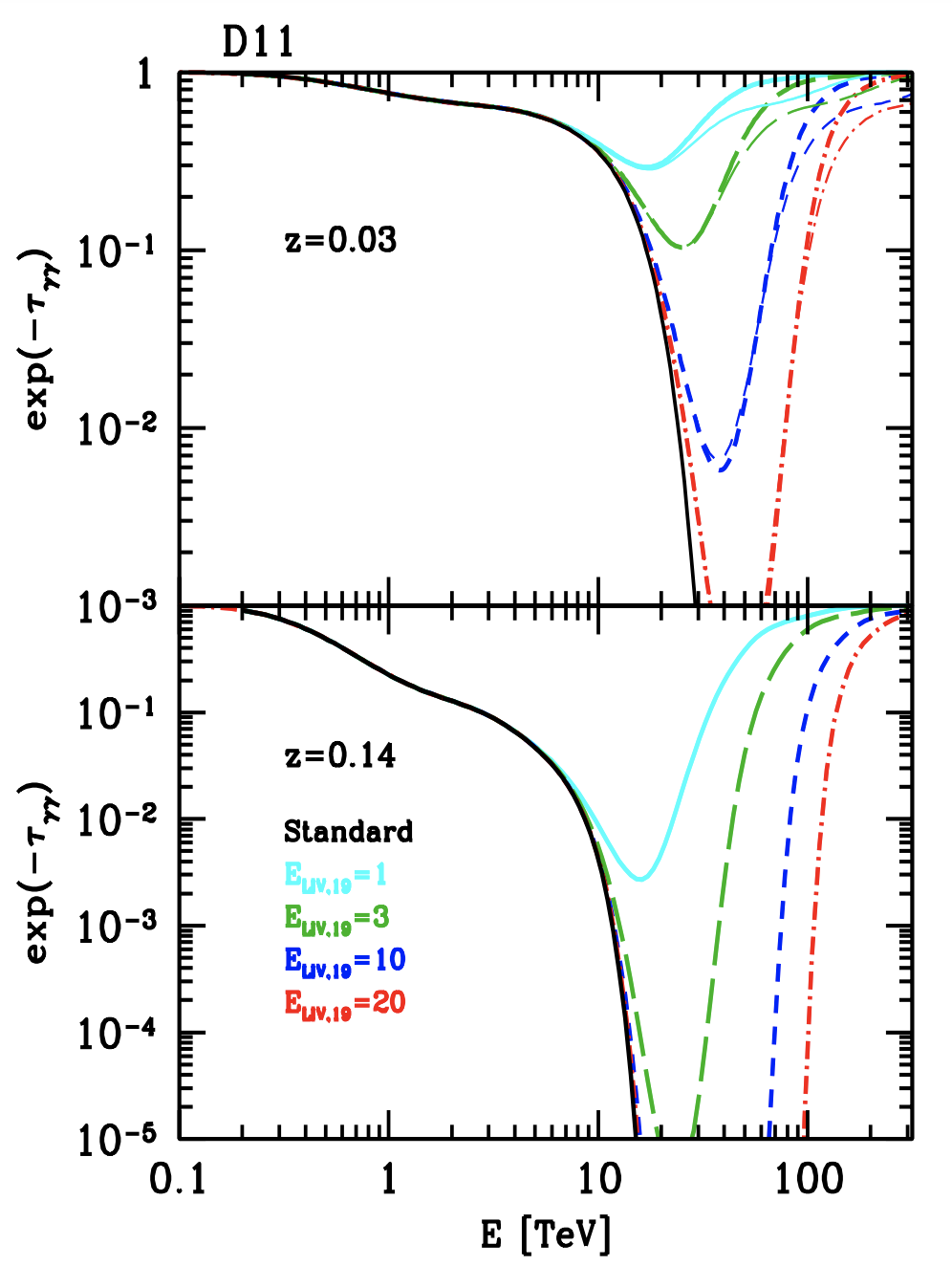}
    \caption{\textit{Top:}Absorption coefficient as a function of energy for gamma rays propagating from a source at $z = 0.03$ \textit{Bottom:} Absorption coefficient as a function of energy for gamma rays propagating from a source at $z = 0.14$. The black solid line refers to the standard case.  From \cite{Tavecchio:2015rfa}}
    \label{fig:Attenuation}
\end{figure}

The LIV superluminal scenario has not been experimentally tested because it is difficult to say whether the more substantial attenuation occurs because of superluminal photons or the source intrinsic effect. As shown in Figure~\ref{fig:Threshold_SED}, in both Lorentz invariant and the LIV superluminal scenario, once the reaction energy threshold is reached, pair production is kinematically allowed for every gamma ray with energy above the reaction energy threshold. This is not the case for the LIV subluminal scenario. In Figure~\ref{fig:Threshold_SED} we can see that the reaction energy threshold increases in the LIV subluminal scenario. Moreover, it is no longer a monotonous function of VHE gamma-ray energy but instead has a global minimum (solid lines in Figure~\ref{fig:Threshold_SED}) and two energy reaction thresholds. Once the second reaction energy threshold is reached, pair creation becomes kinematically forbidden for gamma rays with higher energies. As a result, the opacity of the universe to VHE gamma rays is reduced (Figure~\ref{fig:Attenuation}). Now, a certain number of gamma rays, which would interact with background photons in the Lorentz invariance scenario, now evade absorption and reach our detectors. As a consequence, in the LIV subluminal scenario, we expect a hole in our measured spectra. Namely, photons are first absorbed (flux is suppressed), and then they reemerge as their energy is above the upper reaction energy threshold; thus, they do not interact with background photon fields (for a detailed review, see \cite{Terzic:2021rlx}). This reemergence happens only, of course, if the emitter spectrum extends to higher energies, up to tens of TeV. 
It is hard to do it with the extreme blazars because they are too faint. Another possibility is to use nearby sources. 

\subsection{Time of flight}

Historically, the first alternative scenario for gamma ray propagation that has been tested and where all this began was \textit{Time of flight}. Here, the photon group velocity is no longer $c$ (in vacuum), but depending on S (in Eq. \eqref{eq:Photon_Group_velocity}) will be greater (superluminal) or smaller (subluminal):
\begin{equation}  \label{eq:Photon_Group_velocity}
v_{\gamma}=\frac{\partial E}{\partial p} \simeq\left[1-\sum_{n=1}^{\infty} S \frac{n+1}{2}\left(\frac{E}{E_{\mathrm{QG}, \mathrm{n}}}\right)^{n}\right]\,.
\end{equation}

This effect, of course, is very small and this is one of the reasons why we cannot test it in the laboratory. However, if we take VHE gamma rays emitted from an astrophysical source that has traveled billions of years to reach Earth, we hope that this effect accumulates enough to allow us to detect it. 
The energy-dependent time delay in the arrival of the photons is given by:
\begin{equation} \label{eq:Time_Delay}
\Delta t=t \cdot \Delta v_{\gamma} \cong S \frac{n+1}{2} \frac{E_{h}^{n}-E_{l}^{n}}{E_{\mathrm{QG}, \mathrm{n}}^{n}} \times D_{n}(z)\,.
\end{equation}

The last parameter in Eq.~\eqref{eq:Time_Delay} is the distance contribution. In all studies, up to date, the distance parameter used was one introduced by Jacob \& Piran in 2008 \cite{Jacob:2008bw}, where the distance was derived from the trajectories of comoving particles. There are other approaches, such as DSR. Eq.~\eqref{eq:Time_Delay} assumes simultaneous emission of photons, but, in reality, we do not know whether these two photons were emitted at the same time. For that reason, we do different analysis techniques, the most popular being the maximum likelihood method.

\begin{figure} [H]
    \centering
    \includegraphics[width=0.6\textwidth]{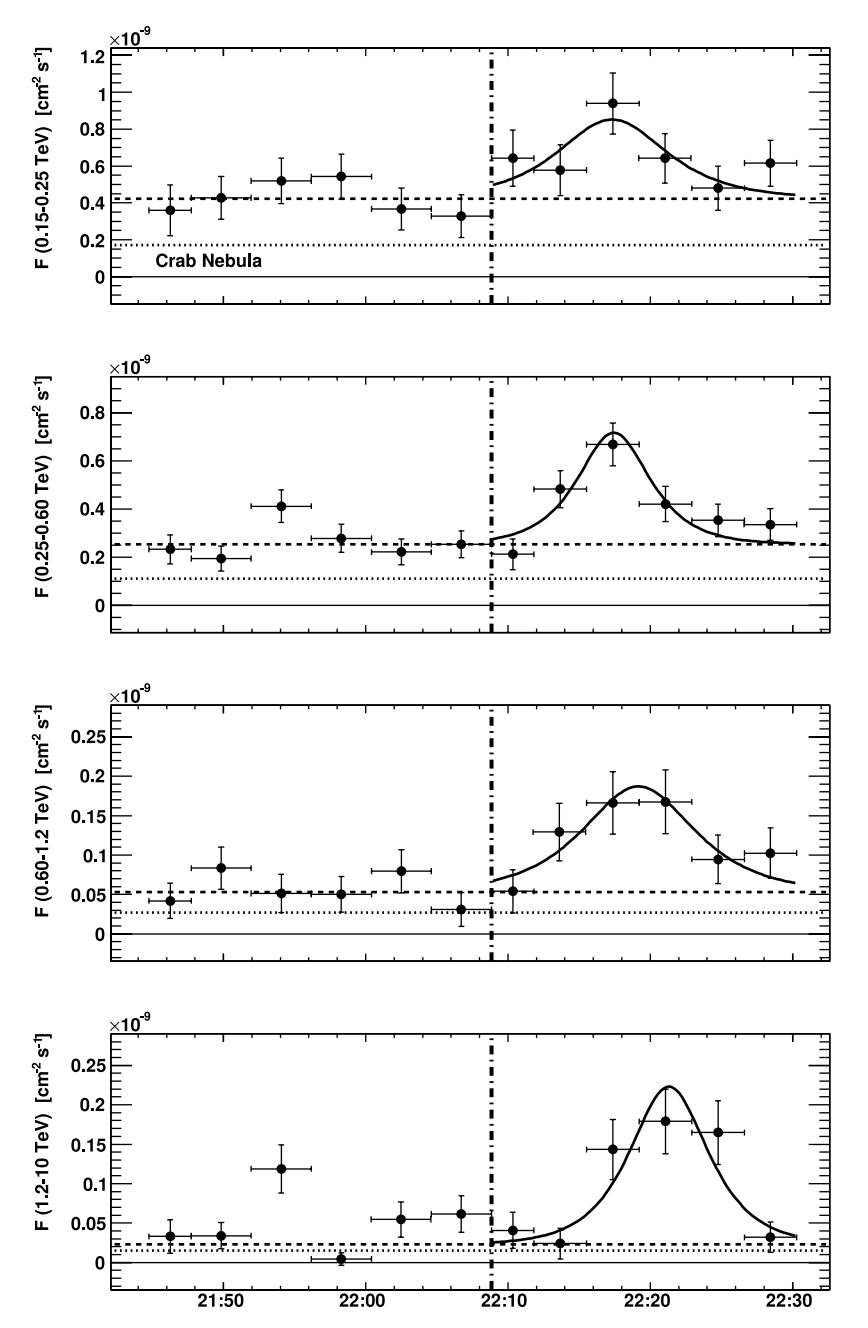}
    \caption{LC of Mrk\,501 for the night July 9 2005. From \cite{2007ApJ...669..862A} }
    \label{fig:LC_Mrk_501}
\end{figure}

If we look at $n=1$ and $n=2$ we have different expectations from the dependence of time variability, distance, and maximum energy detected from a source:  
\begin{equation}
\begin{array}{ccccc}\hline E_{\mathrm{QG}, 1} & \propto & E_{\max } & t_{\mathrm{var}}^{-1} & z_{\mathrm{s}}^{\sim 1} \\ E_{\mathrm{QG}, 2} & \propto & E_{\max } & t_{\mathrm{var}}^{-1/2} & z_{\mathrm{s}}^{\sim 2/3} \\ \hline\end{array}
\end{equation}
In addition, we need good statistics. Here, we pause and make a small remark. Figure~\ref{fig:LC_Mrk_501} represents the LC of a nearby blazar, Mrk\,501 ($z=0.034$). In the summer of 2005, MAGIC registered the fastest flare ever detected from a blazar (still valid) with a flux doubling time of 2 min. The spectrum was extended to 10\,TeV. When only looking at the LC plots, one could conclude that there is some time delay. However, no significant time delay was detected when tested with proper statistical analysis. In this case, two different statistical methods were employed: the energy cost function and maximum likelihood. Therefore, we always need to check our statistics. 

More information on this topic can be found in a review by Terzi\'c et al. \cite{Terzic:2021rlx} and references therein. 

\subsection{Gamma-ALPs oscillation}
Axions are pseudo-Nambu-Goldstone bosons particles proposed to solve the strong CP problem by Peccei and Quin in 1977 \cite{Peccei:1977hh}. Nevertheless, this is not the only physical phenomenon they can (re)solve. Namely, they can decay into two photons, where the decay constant is coupled with axion mass:
\begin{equation}  \label{eq:Axions_Coupling}
g_{A \gamma \gamma}=1 / M\,.
\end{equation}

\begin{figure} [h!]
    \centering
    \includegraphics[width=0.7\textwidth]{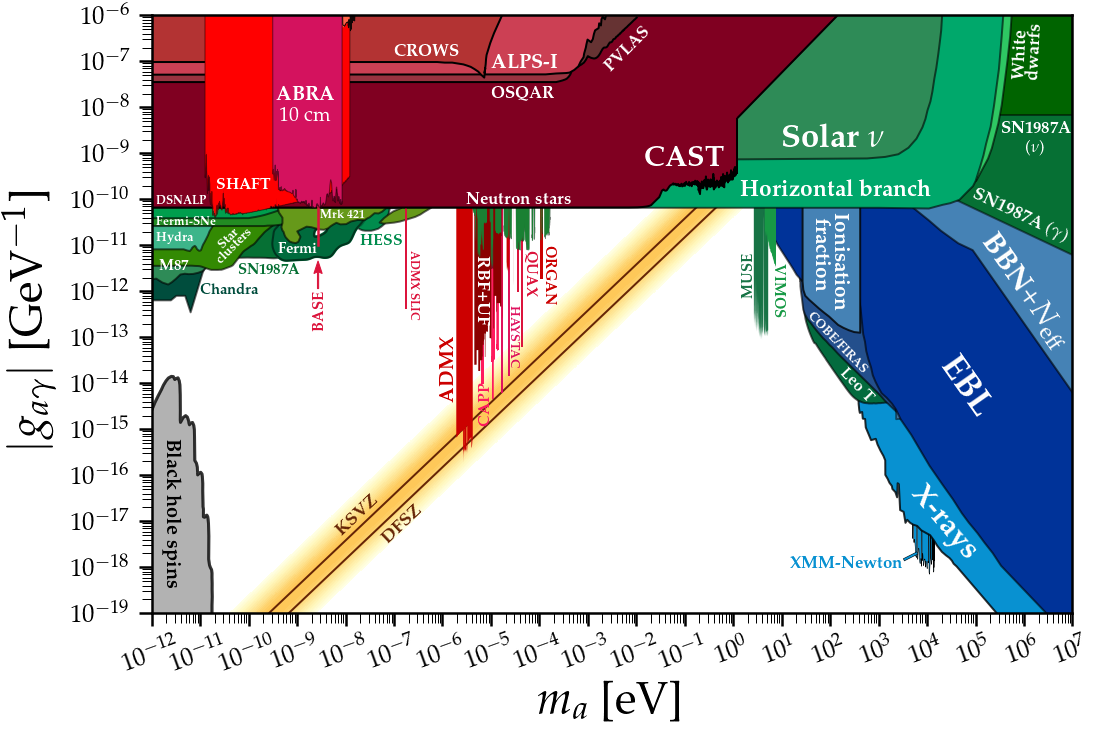}
    \caption{ALPs parameter space with current constraints. The updated limits, references and plots are available in the GitHub repository: https://cajohare.github.io/AxionLimits/}
    \label{fig:ALPs_Parameter}
\end{figure}

\begin{figure} [h!]
    \centering
    \includegraphics[width=0.6\textwidth]{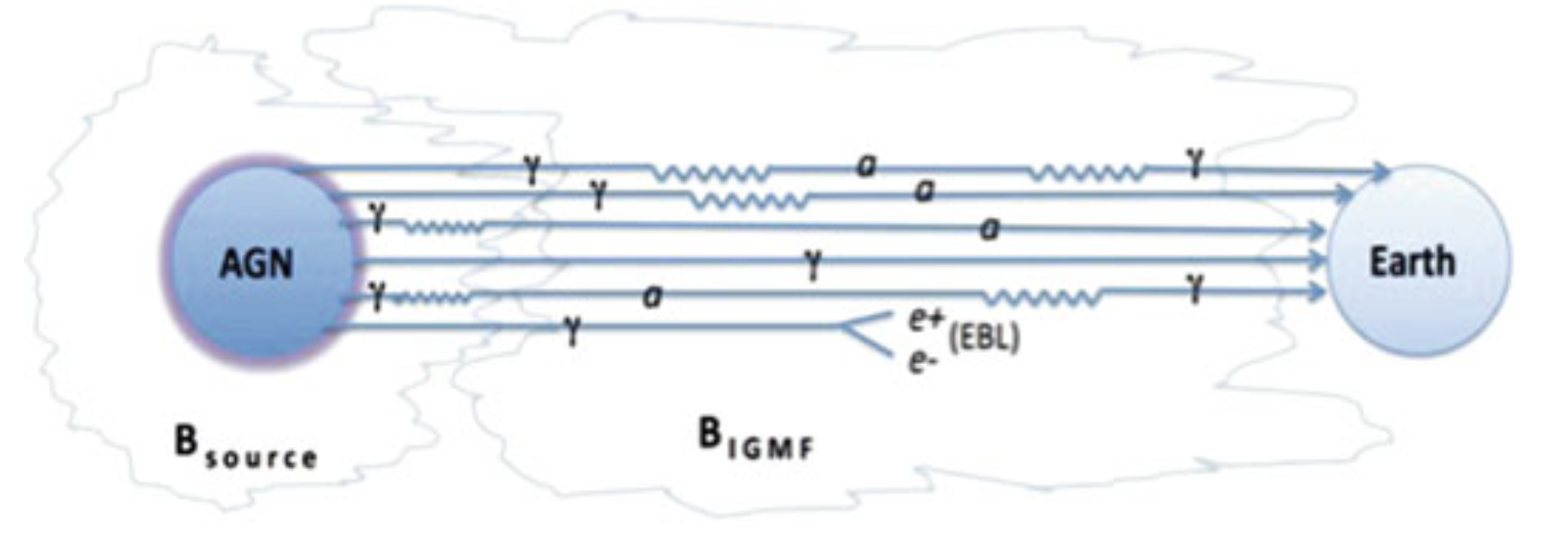}
    \caption{APLs -- gamma-ray oscillations in a presence of a strong magnetic field. }
    \label{fig:ALPs_Gamma:Oscillations}
\end{figure}

Later, the axion model was extended to a whole new group of particles called Axion-Like Particles (ALPs). ALPs are a generalization of axions, but for them, the decay constant is no longer coupled with the axion mass. In addition, ALPs can be found in SM extensions and are proposed as a candidate particle for the dark matter because of their small mass and possibly large decay constant. As such, they play an important role in cosmology. ALPs parameter space with current constraints are depicted in Figure~\ref{fig:ALPs_Parameter}. But, why are we interested in axions? Namely, in strong magnetic fields (e.g., in galaxy clusters), TeV gamma rays could convert into ALPs. Then again, in the presence of a magnetic field of the Milky Way, ALPs can oscillate into photons (Figure~ \ref{fig:ALPs_Gamma:Oscillations}). Since ALPs do not interact with the EBL, we could see more photons than one would expect in standard gamma-ray propagation. In this way, we would see a deviation from standard physics since one does not expect a signal recovery.

The stronger the magnetic field, the stronger the probability of oscillation. To understand current research and the two features mentioned above, we need to describe the concept of critical energy. It depends on the magnetic field and the coupling constant. Below this energy, the probability of photon-ALP mixing is suppressed. The probability becomes sizeable above this energy, and the mixing is maximum. The intermittent photon-ALP-photon conversion could produce an observable effect of longer than expected propagation distances of VHE gamma rays coming from distant TeV sources.
Around this energy, distortions affect gamma-ray spectra. In this case, photon-to-ALP conversion is expected to lead to a detectable energy-dependent distortion of the gamma ray spectra of sources in or behind galaxy clusters. In this scenario, we search for irregularities in our spectrum, often called wiggles. For example, if our expected spectrum from a source is a power law, we look for deviations from this power law. 

\begin{figure} [h!]
    \centering
    \includegraphics[width=0.6\textwidth]{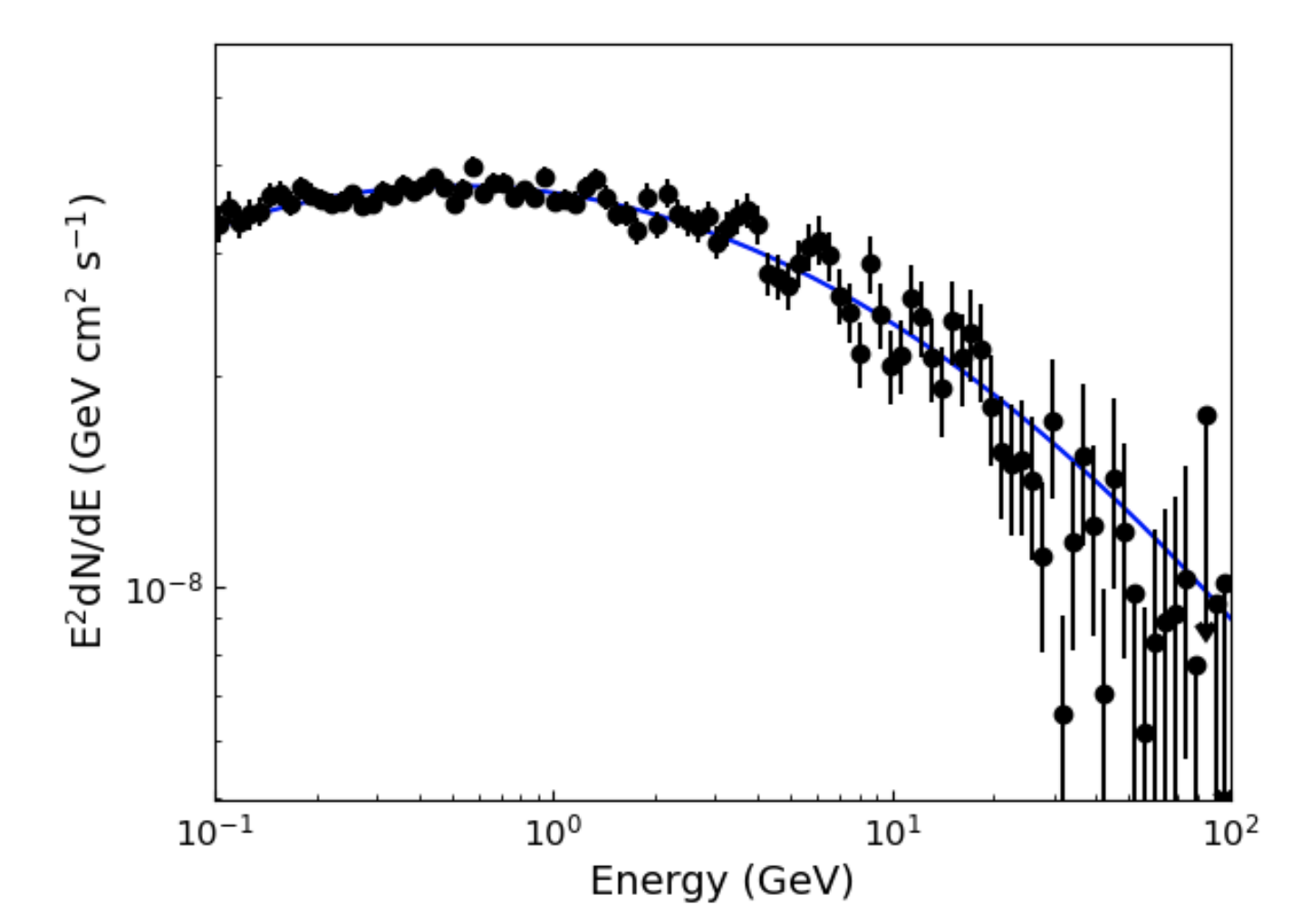}
    \caption{The SED of NGC 1275. It contains 100 energy bins. In each bin, a maximum likelihood analysis is performed to determine the flux and error bar \cite{Cheng:2020bhr}.}
    \label{fig:Fermi_NGC}
\end{figure}

In Figure~\ref{fig:Fermi_NGC} we can see the results obtained by the \textit{Fermi}-LAT collaboration that concentrated its study on the active galaxy NGC~1275. This source is in a galaxy cluster. In galaxy clusters, the magnetic field is much stronger than in extragalactic space, meaning that the probability of ALPs playing a role is much higher.

To summarize, gamma rays can be useful probes to constrain ALPs parameter space. Current limits are based on:
\begin{itemize}
    \item Search for wiggles in the GeV-TeV spectra
    \item Search for anomalies in photon propagation
\end{itemize}
More information on this topic can be found in a review by Batkovi\'c et al. \cite{Batkovic:2021fzr} and references therein. 

\section{Conclusion}
Gamma ray astrophysics is a relatively young and flourishing discipline. Since the early observations, impressive technological advancements have allowed new incredibly performant instruments, both space-born and Earth-based. Understanding propagation effects is becoming more and more crucial in testing cosmological models and theories of physics beyond the standard model. The gamma-ray sky is extremely lively and variable, and many unknowns have to be solved: a challenge for the next generation of instruments and physicists.

\newpage
\section*{Acknowledgements}
The authors thank Tomislav Terzi\'c for his careful review of this manuscript.
EP acknowledges funding from the Italian Ministry of Education, University and Research (MIUR) through the ``Department of Eccellenza'' project Science of the Universe. The work was supported by Nazarbayev University Faculty Development Competitive Research Grant No. 11022021FD2926. This research work was supported by the Hellenic Foundation for Research and Innovation (H.F.R.I.) under the “First Call for H.F.R.I. Research Projects to support Faculty members and Researchers and the procurement of high-cost research equipment grant” (Project Number: 2251). JS acknowledges funding from the University of Rijeka, project number uniri-prirod-18-48. 


\end{document}